\documentclass[12pt]{article}
\usepackage{amsmath}
\usepackage{amssymb}
\usepackage{graphicx}
\usepackage{subfigure}
\usepackage{url}
\begin{document}
\baselineskip 0.6cm

\def\simgt{\mathrel{\lower2.5pt\vbox{\lineskip=0pt\baselineskip=0pt
           \hbox{$>$}\hbox{$\sim$}}}}
\def\simlt{\mathrel{\lower2.5pt\vbox{\lineskip=0pt\baselineskip=0pt
           \hbox{$<$}\hbox{$\sim$}}}}
\def\simprop{\mathrel{\lower3.0pt\vbox{\lineskip=1.0pt\baselineskip=0pt
             \hbox{$\propto$}\hbox{$\sim$}}}}

\begin{titlepage}

\begin{flushright}
MIT-CTP-4402\\
UCB-PTH-12/16
\end{flushright}

\vskip 1.1cm

\begin{center}
{\Large \bf Spread Supersymmetry with {\boldmath $\tilde{W}$} LSP:\\
\vspace{1mm}
Gluino and Dark Matter Signals}

\vskip 0.7cm

{\large Lawrence J. Hall$^a$, Yasunori Nomura$^{a,b}$, and Satoshi Shirai$^a$}

\vskip 0.4cm

$^a$ {\it Berkeley Center for Theoretical Physics, Department of Physics, \\
     and Theoretical Physics Group, Lawrence Berkeley National Laboratory, \\
     University of California, Berkeley, CA 94720, USA} \\

\vskip 0.2cm

$^b$ {\it Center for Theoretical Physics, Laboratory for Nuclear Science, 
     and Department of Physics, \\
     Massachusetts Institute of Technology, Cambridge, MA 02139, USA} \\

\vskip 0.8cm

\abstract{The discovery of a Higgs boson near $125~{\rm GeV}$, together 
 with the absence of LHC signals for supersymmetry or direct detection 
 signals of dark matter, motivate further study of a particular theory 
 of split supersymmetry.  In arguably the theoretically simplest 
 implementation of split, the superpartner spectrum is spread over 
 several decades.  The squarks and sleptons are heavier than the gravitino 
 and Higgsinos by a factor $M_{\rm Pl}/M_*$, where $M_*$ is the mediation 
 scale of supersymmetry breaking and is high, between unified and Planck 
 scales.  On the other hand the gaugino masses are 1-loop smaller than 
 the gravitino and Higgsino masses, arising from both anomaly mediation 
 and a Higgsino loop.  Wino dark matter arises from three sources:\ 
 gravitino production by scattering at high temperatures, gravitino 
 production from squark decays, and thermal freeze-out.  For reheating 
 temperatures larger than the squark mass, these conspire to require 
 that the squarks are lighter than about $10^4~{\rm TeV}$, while collider 
 limits on gaugino masses require squarks to be heavier than about 
 $100~{\rm TeV}$.  Whether winos constitute all or just a fraction of 
 the dark matter, a large fraction of the allowed parameter space has 
 the gluino within reach of the LHC with $0.1~{\rm mm} < c\tau_{\tilde{g}} 
 < 10~{\rm cm}$, leading to displaced vertices.  In addition, events 
 with cascades via $\tilde{W}^\pm$ lead to disappearing charged tracks 
 with $c\tau_{\tilde{W}^\pm} \sim 10~{\rm cm}$.  The squarks and sleptons 
 are predicted to be just heavy enough to solve the supersymmetric flavor 
 and $CP$ problems.  Thus gluino decay modes may typically violate flavor 
 and involve heavy quarks: $[ \bar{t}(t,c,u)+\bar{b}(b,s,d)] \tilde{W}^0$ 
 and $[ \bar{t}(b,s,d)+(\bar{t},\bar{c},\bar{u})b] \tilde{W}^\pm$. 
 The electron electric dipole moment is expected to be of order 
 $10^{-29}~e\,{\rm cm}$, two orders of magnitude below the current 
 limit.  The AMS-02 search for cosmic ray antiprotons will probe 
 an interesting region of parameter space.}

\end{center}
\end{titlepage}

\section{Introduction}
\label{sec:intro}

All realistic theories of supersymmetry have supersymmetry breaking in 
a hidden sector---the key question is how this breaking is mediated to 
the superpartners of the standard model (SM) particles.  In 4 dimensions, 
physics at the gravitational scale provides an almost irremovable 
contribution to the mediation~\cite{Chamseddine:1982jx}, and generically 
leads to all SM superpartners acquiring masses of order the gravitino 
mass, $m_{3/2} = F_X/\sqrt{3} M_{\rm Pl}$, where $F_X$ is the leading 
spurion of supersymmetry breaking and $M_{\rm Pl}$ is the reduced Planck 
scale.  Augmented with an approximate flavor symmetry, gravity mediation 
could describe supersymmetry breaking with few parameters, leading 
to theories where dark matter arises from freeze-out of the lightest 
supersymmetric particle (LSP).

Over the last 30~years, this minimal standard picture of gravity mediation 
met a succession of challenges:

{\bf The Supersymmetric Flavor/{\boldmath $CP$} Problem} arises because 
it is not clear that approximate flavor symmetries will be respected 
by physics at the gravitational scale.  However, solutions such as gauge 
mediation require further fields and model building, and do not allow 
weakly interacting massive particle (WIMP) LSP dark matter in their 
minimal implementations.

{\bf A Derived Planck scale:} With extra spatial dimensions, the Planck 
scale is a derived scale, larger than the fundamental scale $M_*$ by 
a volume factor~\cite{Horava:1995qa}.  In this case higher dimensional 
operators arise in the low energy effective theory from integrating 
out string states, leading to supersymmetry breaking masses of order 
$\tilde{m} = F_X/M_*$, dominating the purely gravitational ones. 
This typically yields a gravitino LSP, again precluding WIMP LSP dark 
matter.  Furthermore, decays of the next-to-LSP to the gravitino in 
the early universe occur after nucleosynthesis and are generically 
problematic~\cite{Khlopov:1984pf,Kawasaki:2008qe}.

{\bf Anomaly Mediation:} A hidden assumption of gravity mediation is that 
the field $X$ of the supersymmetry breaking sector is neutral under all 
symmetries.  If it is charged under some symmetry, operators linear in 
$X$ leading to gaugino masses and the $\mu$ parameter are absent, so that 
the leading supersymmetry breaking in the SM sector is for scalar mass 
terms only.  The dominant contribution to gaugino masses arises at 1-loop 
via the superconformal anomaly~\cite{Randall:1998uk,Giudice:1998xp}, yielding 
an $O(\alpha/4\pi)$ hierarchy between scalar and fermion superpartner 
masses that destroys the supersymmetric solution to the hierarchy problem. 
One can either attempt to regain naturalness by suppressing the contribution 
from gravity mediation by sequestering~\cite{Randall:1998uk}, or one 
can accept that the theory possesses a few orders of magnitude of 
fine-tuning~\cite{Giudice:1998xp,Wells:2003tf}.

{\bf Split Supersymmetry:} The environmental requirement of structure 
formation allows a multiverse solution to the cosmological constant 
problem~\cite{Weinberg:1987dv,Martel:1997vi}; similarly, the requirement 
of stable complex nuclei allows a multiverse solution to the hierarchy 
problem~\cite{Agrawal:1997gf}.  Furthermore, both environmental arguments 
become plausible in the context of the string landscape~\cite{Bousso:2000xa}. 
While a multiverse solution to the hierarchy problem decouples the scale 
of supersymmetry breaking from the weak scale, the fermionic superpartners 
have chiral symmetries that could allow them to be much lighter than the 
scalar superpartners to account for dark matter, yielding a highly split 
spectrum of superpartners~\cite{ArkaniHamed:2004fb}.  A variety of split 
spectra yield gauge coupling unification that is as precise as natural 
supersymmetric theories.

How does the discovery of a Higgs boson near $125~{\rm GeV}$, and the 
absence of signals for supersymmetry so far, affect our view of the 
mediation of supersymmetry breaking?  Certainly there is no unique 
answer---at one extreme high scale supersymmetry~\cite{Hall:2009nd}, 
with the SM valid to unified scales, remains a possibility if $\tan\beta$ 
is close to unity, and at the other extreme there are several possibilities 
for natural electroweak symmetry breaking that allow superpartners to 
evade current LHC searches~\cite{Papucci:2011wy}.  However, there is 
a very simple scenario that addresses all four of the above challenges, 
is well-motivated by LHC results to date, and has highly distinctive 
LHC and astrophysical signals.

The first key assumption is that the field $X$ carries some symmetry so that,
from the above discussion, the superpartner spectrum has a modest degree 
of splitting from two sources:
\begin{itemize}
\item Gaugino masses arising from anomaly mediation are $O(\alpha/4\pi)$ 
suppressed relative to scalar superpartner masses.
\item Gaugino and Higgsino masses arising only from gravitational effects 
are $O(M_*/M_{\rm Pl})$ suppressed relative to scalar superpartner masses.
\end{itemize}
If these are the only sources of splitting, the spectrum of the ``Simplest 
Model of Split Supersymmetry"~\cite{ArkaniHamed:2006mb} results---the 
case we study in this paper.

With this moderately split spectrum, the weak scale is fine-tuned by several 
orders of magnitude, making plausible the second key assumption:
\begin{itemize}
\item The overall normalization of the superpartner spectrum is determined 
by an environmental requirement on the abundance of dark matter.
\end{itemize}
This selection in the multiverse (or quantum many 
universes~\cite{Nomura:2011dt}) has an important implication---dark 
matter might be multi-component, for example with roughly comparable 
contributions from LSPs and axions.  This will have the effect of 
increasing the range of parameters that yields signals at the LHC. 
Taken together, the above three items define what we mean by ``Spread 
Supersymmetry''~\cite{Hall:2011jd}.

If $X$ is neutral under all symmetries, differentiating $M_*$ from 
$M_{\rm Pl}$ leads to the expectation of a gravitino LSP.  However, 
taking $X$ charged leads instead to either a gaugino or Higgsino LSP. 
The origin of the Higgsino mass is critical since it determines whether 
the LSP is gaugino or Higgsino, leading to two realistic versions of 
Spread Supersymmetry with neutral wino or Higgsino dark matter.  In 
this paper we assume the Higgsino mass is of order $m_{3/2}$ so that 
the LSP is wino, since only in this case are we led to interesting 
gluino signals at the LHC.  This can arise from supergravity 
interactions that follow from having $H_u H_d$ in the K\"{a}hler 
potential~\cite{Giudice:1988yz} or that cause a readjustment of 
the vacuum~\cite{Hall-mu,Hempfling:1994ae}.

\begin{figure}[t]
\begin{center}
  \scalebox{1}{
\begin{picture}(220,195)(95,-7)

\put(166,164.014){$\scriptstyle \tilde{q}, \tilde{l} , H$}
\put(160,160.014){\line(1,0){30}}
\put(160,159.014){\line(1,0){30}}
\put(160,158.014){\line(1,0){30}}
\put(174,153.014){$\cdot$}
\put(174,151.014){$\cdot$}
\put(174,149.014){$\cdot$}
\put(160,150.514){\line(1,0){30}}
\put(172,125){$\scriptstyle \tilde{G}$}
\put(160, 122){\line(1,0){30}}
\put(160, 110){\line(1,0){30}}
\put(173,100){$\scriptstyle \tilde{h}$}
\put(174,46){$\scriptstyle \tilde{g}$}
\put(160,42){\line(1,0){30}}
\put(171,28.5){$\scriptstyle  \tilde B$}
\put(160,25.5){\line(1,0){30}}
\put(171,11.5){$\scriptstyle \tilde W$}
\put(160,19){\line(1,0){30}}
\put(160, 9){\line(1,0){30}}
\put(171,0.5){$\scriptstyle h^0$}

\put(250,86.4404){\small Mass}
\put(252.5,76.4404){$\scriptstyle {\rm [TeV]}$}
\put(220,-10){\vector(0,1){194}}
\put(220,0){\line(1,0){10}}
\put(232,-2){$\scriptstyle 0.1$}
\put(220,40){\line(1,0){10}}
\put(232,38){$\scriptstyle 1$}
\put(220,80){\line(1,0){10}}
\put(232,78){$\scriptstyle 10$}
\put(220,120){\line(1,0){10}}
\put(232,118){$\scriptstyle 100$}
\put(220,160){\line(1,0){10}}
\put(232,158){$\scriptstyle 1000$}

\end{picture} }
\caption{Typical spectrum of Spread Supersymmetry with wino LSP.}
\label{fig:spread}
\end{center}
\end{figure}
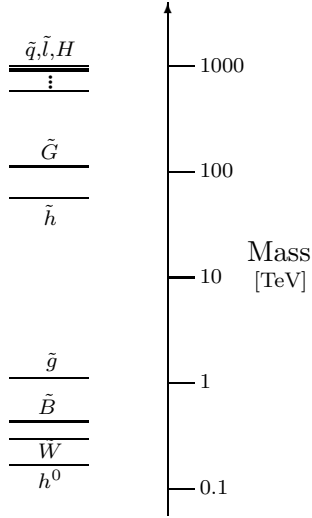

Hence, we study a supersymmetric theory with minimal field content and 
leading supersymmetry breaking effects arising from
\begin{equation}
  {\cal L}_{\rm SB} \,\sim\, \frac{1}{M_*^2}\, [X^\dagger X\, 
      (\Phi^\dagger \Phi  + H_u H_d)]_{\theta^4} 
    + [H_u H_d]_{\theta^4} - \frac{m_{3/2}}{2} \left( \tilde{G}\tilde{G} 
      + \frac{b_a\alpha_a}{4\pi}\, \lambda_a \lambda_a \right),
\label{eq:susybr}
\end{equation}
where $\Phi$ are the chiral superfields, $H_u, H_d$ are the Higgs superfields, 
$\tilde{G}$ is the gravitino, $\lambda_a$ are the gauginos, and $b_a$ 
and $\alpha_a$ are the 1-loop beta function coefficients and gauge coupling 
strengths for $a = U(1)_Y, SU(2)_L, SU(3)_C$.  Here, we have omitted the 
chiral compensator field $\phi = 1 + m_{3/2} \theta^2$, which is important 
in the second term leading to the supersymmetric Higgs mass $\mu$ of 
order $m_{3/2}$, and each operator (except for the $\tilde{G}$ and 
$\lambda_a$ mass terms) should be understood to have an unknown coefficient 
of order unity that is not displayed.  The resulting superpartner masses 
have a moderate hierarchy
\begin{equation}
  (\tilde{q}, \tilde{l}, H) \,\,:\,\, (\tilde{G}, \tilde{h}) \,\,:\,\, 
    \lambda_a \,\,\,\,\approx\,\,\,\, \tilde{m} \,\,:\,\, m_{3/2} 
    \,\,:\,\, \frac{\alpha_a}{4\pi} m_{3/2},
\label{eq:susyspectrum}
\end{equation}
as depicted in Fig.~\ref{fig:spread}.  Here $H$ is the heavy Higgs doublet 
orthogonal to the doublet that has been fine-tuned to be at the weak scale.

Many aspects of a moderately split supersymmetric spectrum have been noted 
and studied before.  In particular, anomaly mediation triggered the first 
papers to take seriously the unpopular idea of a few orders of magnitude 
of fine-tuning~\cite{Giudice:1998xp,Wells:2003tf}, and also triggered studies 
of wino dark matter~\cite{Gherghetta:1999sw,Hisano:2005ec,Hisano:2010ct,%
Hisano:2004ds}.  The possibility of combining this with a modest hierarchy 
from $M_*/M_{\rm Pl}$ to yield a solution to the flavor problem was mentioned 
in~\cite{ArkaniHamed:2004fb}, together with the possibility of long-lived 
gluinos.  Implications of this scenario on thermal WIMP dark matter, 
especially the possibility of having dark matter with significant mass 
degeneracy between a bino and wino, were studied in~\cite{ArkaniHamed:2006mb}. 
It is well-known that once the gravitino mass is above $10~{\rm TeV}$ 
there is no cosmological gravitino problem, and similarly a moderately 
split spectrum solves possible moduli and proton decay problems.  LHC 
signatures of wino LSP were studied in~\cite{Ibe:2006de,Asai:2007sw}, 
and aspects of particle physics and cosmology of a moderately split 
spectrum with $M_* = M_{\rm Pl}$ were discussed recently in a series 
of papers in~\cite{Ibe:2011aa}.

\begin{figure}[t]
\begin{center}
  \includegraphics[clip,width=0.65\textwidth]{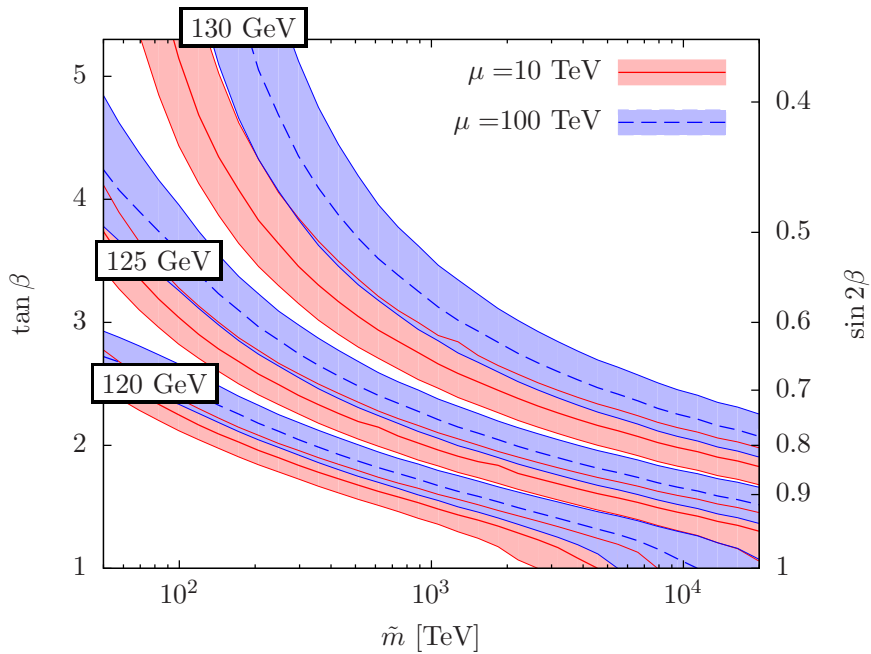}
\caption{The values of the Higgs boson mass in the $\tilde{m}$-$\tan\beta$ 
 plane.  The solid (red) curves represent ones with $\mu = 10~{\rm TeV}$, 
 while the dashed (blue) curves $\mu = 100~{\rm TeV}$.  The shaded region 
 around each curve shows uncertainty from the top quark mass.  For the 
 gaugino masses, we have set $M_1 = 600~{\rm GeV}$, $M_2 = 300~{\rm GeV}$, 
 and $M_3 = 2000~{\rm GeV}$.}
\label{fig:higgsmass}
\end{center}
\end{figure}

Dark matter is a critical aspect of Spread Supersymmetry since it determines 
the normalization of the entire superpartner spectrum.  It constrains the 
scale of the squark masses to be in the range
\begin{equation}
  \tilde{m} \sim (10^2-10^4)~{\rm TeV}.
\label{eq:mtilde}
\end{equation}
The upper limit follows from freeze-in of dark matter via 
gravitinos~\cite{ArkaniHamed:2004yi,Hall:2009bx}, assuming $T_R > 
\tilde{m}$, and the lower limit from requiring gravitinos to decay 
before the Big-Bang nucleosynthesis (BBN).  Furthermore, from 
Eq.~(\ref{eq:susybr}) all entries in the Higgs mass-squared matrix 
are comparable, so that $\tan\beta$ is not expected to be large.  Since 
the top squark mixing parameter vanishes at tree level, the Higgs boson 
mass is determined essentially only by $(\tilde{m}, \tan\beta)$, which 
is shown in Fig.~\ref{fig:higgsmass}.  The measurement of the Higgs 
mass at the LHC is a key motivation for studying the predictions of 
this theory in some detail.

In theories with wino LSP arising from anomaly mediation, it is generally 
understood that the wino mass should be near $3~{\rm TeV}$, so that thermal 
freeze-out can account for the observed dark matter, leading to a gluino 
heavier than $5~{\rm TeV}$ that is out of the LHC reach.  In this paper 
we show that in Spread Supersymmetry the situation is different:\ there 
is a large fraction of parameter space where the gluino is light enough 
to be discovered at the LHC.  There are two reasons for this reduction 
in the wino and gluino masses.  Firstly dark matter production also occurs 
via gravitino production, from both freeze-in from squark decay and UV 
gluino scattering.  Secondly, the winos might account for only a fraction 
of the observed dark matter, as discussed below.  Both these effects 
require a lower wino mass.  Another consequence of the lighter wino 
is the change in the indirect cosmic ray signals for dark matter.  The 
photon signal already excludes winos lighter than about $500~{\rm GeV}$ 
constituting all the dark matter, and antiproton signals will soon provide 
an additional probe of the theory.

The lifetime of the gluino reaches $c \tau_{\tilde{g}} \approx 10~{\rm m}$, 
for $\tilde{m}$ at the upper limit of Eq.~(\ref{eq:mtilde}) for a 
$1~{\rm TeV}$ gluino mass, suggesting an exciting possibility of displaced 
vertices from gluino decays~\cite{ArkaniHamed:2004fb}.  However, the 
lifetime scales as $\tilde{m}^4$ and hence a detailed analysis of dark 
matter production, including freeze-out, freeze-in, and UV scattering, 
is required to predict the allowed range of $c\tau_{\tilde{g}}$.  Such 
an analysis will also yield the precise allowed range of $\tilde{m}$, 
estimated at the order of magnitude level in Eq.~(\ref{eq:mtilde}), giving 
an indication of the likelihood of flavor and $CP$ violating signals 
in this theory.

In the context of the multiverse, the abundance of dark matter can serve 
an important environmental factor that affects the selection of parameters 
of the theory in our universe.  In particular, this may choose the initial 
misalignment angle of the axion field---which we expect to exist because 
of the strong $CP$ problem---yielding axion dark matter~\cite{Linde:1987bx}. 
In Spread Supersymmetry, the same mechanism can act on the combined 
abundance of relic WIMPs and axions, leading generically to multi-component 
dark matter.  The precise ratio of the two components depends on the 
a priori distribution of parameters in the landscape~\cite{Hall:2011jd}. 
We, however, expect in general that the LSP abundance is {\it bounded} by 
the observed dark matter abundance, $\Omega_{\tilde{W}} < \Omega_{\rm DM}$, 
not necessarily saturates it, although how much $\Omega_{\tilde{W}}$ deviates 
from $\Omega_{\rm DM}$ depends on the a priori probability distribution 
of parameters.

The organization of this paper is as follows.  In Section~\ref{sec:particle}, 
we study particle physics aspects of the model.  We describe the detailed 
spectrum of superpartners and the Higgs boson, and analyze collider 
signals as well as physics of flavor and $CP$ violation.  We find that 
the model naturally leads to a distinct signal of a long-lived gluino 
decaying into a long-lived charged wino, decaying into the neutral wino 
LSP.  The model also allows for interesting handles of the flavor structure 
in the scalar sector at $\tilde{m} \sim (10^2-10^4)~{\rm TeV}$ through 
observations of gluino decays.  In Section~\ref{sec:cosmo}, we study 
astrophysical and cosmological aspects of the model.  We find that, 
because of relatively large $\tilde{m}$, the freeze-in contribution 
to $\Omega_{\tilde{W}}$ is generally important.  We also consider the 
UV scattering and thermal contributions, and discuss their effects on 
the cosmology of the model.  In Section~\ref{sec:results}, we combine 
these results and study current constraints on the model as well as 
future prospects for a discovery.  We find that while some of the parameter 
space is already constrained by the current LHC and Fermi data, there are 
large parameter regions still unconstrained, some of which is compatible 
with thermal leptogenesis at high temperatures.  We find that future 
data from the LHC, searches of electric dipole moments, and astrophysical 
observations have good potentials to discover signals of the model. 
In Section~\ref{sec:multiverse}, we discuss how the current model may 
arise from environmental selection in the multiverse.  Finally, we 
conclude in Section~\ref{sec:concl}.

\section{Particle Physics}
\label{sec:particle}

In this section we discuss particle physics aspects of Spread Supersymmetry 
with wino LSP.  We first describe the detailed spectrum of the theory, 
and then discuss physics at colliders and of flavor and $CP$ violations.

\subsection{Mass spectrum}
\label{subsec:spectrum}

We assume that the supersymmetry breaking field $X$ is charged under some 
symmetry.  While this suppresses the operators linear in $X$, it still 
allows $X^\dagger X$ to couple to any SM singlet operators; in particular, 
it allows the K\"{a}hler potential terms
\begin{equation}
  K \ni -\frac{c}{M_*^2} X^\dagger X \Phi_{\rm MSSM}^\dagger \Phi_{\rm MSSM},
\label{eq:K-scalar}
\end{equation}
where $\Phi_{\rm MSSM} = \Phi_M, H_u, H_d$, and $c = O(1)$ represents 
generic coefficients, which may depend on the species.  This leads to 
the soft masses $\tilde{m}^2 = c|F_X|^2/M_*^2$ for the minimal supersymmetric 
standard model (MSSM) scalars.  Here, $M_*$ is the cutoff scale of the 
theory, which we take to be larger than the supersymmetric unification 
scale, $\approx 10^{16}~{\rm GeV}$, to preserve supersymmetric gauge 
coupling unification in its simplest form.%
\footnote{If $M_*$ is too close to the unification scale, we may expect 
 significant threshold corrections from higher dimension operators.}

We expect scale $M_*$ to be smaller than the 4D reduced Planck scale, 
$M_{\rm Pl}$.  For example, if there is a small extra $d$-dimensional 
space around these scales, then we have $M_{\rm Pl}^2 \approx M_*^{2+d} 
V > M_*^2$, where $V$ is the volume of the extra compact space; more 
generally, if there are $N \gg 1$ species that are effectively massless 
around these scales, then we expect $M_{\rm Pl}^2 \approx N M_*^2 > 
M_*^2$~\cite{Dvali:2007hz}.  This implies that the scalar masses 
$\tilde{m} \approx F_X/M_*$ is expected to be somewhat larger than 
the gravitino mass $m_{3/2} = F_X/\sqrt{3}M_{\rm Pl}$.  For later 
convenience, we define
\begin{equation}
  r_* \equiv \frac{\sqrt{3} M_{\rm Pl}}{M_*} 
  \approx \frac{\tilde{m}}{m_{3/2}}.
\label{eq:r_*}
\end{equation}
In this paper we mainly consider the range $1 \simlt r_* \simlt O(100)$.

Let us now discuss the Higgs sector.  If $H_u H_d$ is neutral, which we 
assume here, then we can have operators
\begin{equation}
  K \ni -\frac{c'}{M_*^2} X^{\dagger} X H_u H_d + c'' H_u H_d + {\rm h.c.},
\end{equation}
which generate the holomorphic supersymmetry-breaking Higgs mass-squared 
$b = c'|F_X|^2/M_*^2 $ as well as the supersymmetric Higgs mass $\mu 
= c'' m_{3/2}^*$.  Unlike the soft scalar masses, the supersymmetric 
$\mu$ term is of order the gravitino mass.%
\footnote{We assume there is no superpotential term of the form $W = 
 H_u H_d \langle W \rangle/M_*^2$, where $\langle W \rangle$ is the 
 expectation value of the superpotential needed to cancel the cosmological 
 constant, which would lead to $\mu = O(m_{3/2} M_{\rm Pl}^2/M_*^2)$.}
The size of the $b$ term is of the same order as the soft scalar mass-squared, 
which implies that $\tan\beta \equiv \langle H_u \rangle/\langle H_d 
\rangle= O(1)$.

Direct couplings of $X$ to the gauge supermultiplets are forbidden by 
the symmetry, so that the main contribution to the gaugino masses arise 
from anomaly mediation.  In addition, the electroweak gauginos obtain 
masses of comparable size from loops of the Higgsino and Higgs bosons. 
The gaugino mass parameters are then given by
\begin{eqnarray}
  M_1 &=& \frac{3}{5} \frac{\alpha_1}{4\pi} (11 m_{3/2} + L),
\label{eq:M1}\\
  M_2 &=& \frac{\alpha_2}{4\pi} (m_{3/2} + L),
\label{eq:M2}\\
  M_3 &=& \frac{\alpha_3}{4\pi} (-3 m_{3/2})(1 + c_{\tilde{g}})
\label{eq:M3}.
\end{eqnarray}
Here, $L$ represents the correction from Higgsino-Higgs loops:
\begin{equation}
  L = \mu \sin(2\beta) \frac{m_A^2}{|\mu|^2 - m_A^2} 
    \ln\frac{|\mu|^2}{m_A^2}
  \sim 2\mu \sin(2\beta) \ln r_*,
\label{eq:L}
\end{equation}
with $m_A = O(\tilde{m})$ being the heavy Higgs boson mass. 
$c_{\tilde{g}}$ is the logarithmic correction from the heavy squarks, 
which at the one-loop level is given by
\begin{equation}
  1 + c_{\tilde{g}} = \left( 1 + \frac{5 \alpha_3(|M_3|)}{4\pi} 
    \ln\frac{m_{\tilde{q}}}{|M_3|} \right)^{4/5},
\end{equation}
with $m_{\tilde{q}} = O(\tilde{m})$ being the squark mass which we 
have taken to be universal here.  In the expressions above and throughout 
the paper, we adopt the phase convention that the gravitino mass $m_{3/2}$ 
and $\tan\beta$ are real and positive, while $\mu$ has a complex phase 
in general.  Note that we can always take this convention by appropriate 
phase rotations of the fields.

The gauginos are lightest among all the superparticles.  The condition 
$m_{\rm gaugino} \simgt 100~{\rm GeV}$ then indicates $m_{3/2} > 
O(10~{\rm TeV})$, so that the scalar masses are $\tilde{m} \approx 
r_* m_{3/2}$, which are typically of order $10^2~\mbox{--}~10^4~{\rm TeV}$. 
Such heavy scalars predict a relatively large SM-like Higgs boson mass 
$m_h$.  In fact, we find that the suggested values of $\tan\beta = O(1)$ 
and $\tilde{m} = O(10^2~\mbox{--}~10^4~{\rm TeV})$ naturally realize the 
$125~{\rm GeV}$ Higgs boson as seen in Fig.~\ref{fig:higgsmass}.  (Note 
that scalar trilinear interactions are generated only by anomaly mediation 
and thus are small $A = O(m_{3/2}/16\pi^2) \ll \tilde{m}$.)

Phenomenology of the model depends strongly on which gaugino is the LSP. 
As seen in Eqs.~(\ref{eq:M1}~--~\ref{eq:M3}), the relative values of the 
gaugino masses depend on the size of $L$.  Since $L$ can be relatively 
large in the present model because of large $r_*$, in principle any 
gaugino can be the LSP.  However, if $R$-parity conservation is assumed, 
the LSP is stable and contributes to the dark matter.  In this case the 
gluino LSP is excluded.  In the case of the bino LSP, its thermal relic 
abundance is roughly given by
\begin{equation}
  \Omega_{\tilde{B}} h^2 \sim O(10^{2-3}) \times 
    \left( \frac{|\mu|}{10~{\rm TeV}} \right)^2
\label{eq:Omega-B}
\end{equation}
for $\tan\beta \sim 1$ and $|M_1| \ll |M_2|$, where $h \simeq 0.7$ is 
the present-day Hubble expansion rate in units of ${\rm km}\, {\rm s}^{-1}\, 
{\rm Mpc}^{-1}$.  The precise value depends on details, especially the 
phase of $M_1 \mu$ since its imaginary part contributes to $s$-wave 
annihilation, but the value suggested by Eq.~(\ref{eq:Omega-B}) is much 
larger than the observed dark matter abundance for a natural parameter 
region, $|\mu| = O(m_{3/2})$.  Therefore, we focus on 
the case of the wino LSP in this paper.%
\footnote{If the bino mass is about $60~{\rm GeV}$ and its annihilation 
 hits the $125~{\rm GeV}$ Higgs boson pole, then the size of $\mu$ 
 can be as large as a few TeV without a $CP$ violating phase, avoiding 
 overabundance.  If $|M_2| - |M_1|$ is small, coannihilation process 
 is effective~\cite{ArkaniHamed:2006mb}, and the bino LSP is allowed. 
 In this case, $(|M_2| - |M_1|)/|M_1| < O(10\%)$ is required.}

\begin{figure}[t]
\begin{center}
  \includegraphics[clip,width=0.65\textwidth]{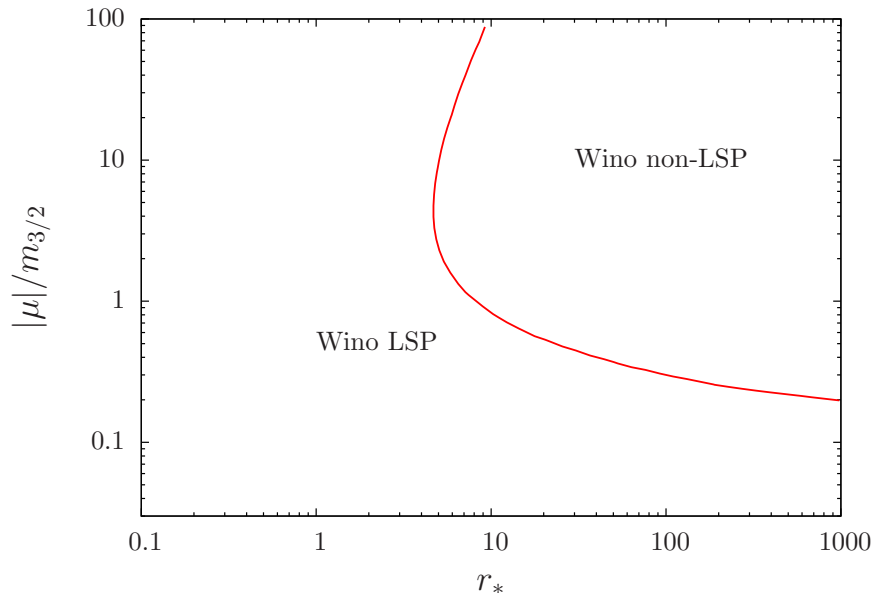}
\caption{The wino LSP region in the $r_*$-$|\mu|/m_{3/2}$ plane.  In the 
 ``Wino LSP'' region, an appropriate choice of $\arg(\mu)$ allows the wino 
 LSP, while in the ``Wino non-LSP'' region, no choice of $\arg(\mu)$ leads 
 to the wino LSP.}
\label{fig:LSP}
\end{center}
\end{figure}

The requirement of the wino LSP constrains the size of $|\mu|$, depending 
on the value of $r_*$, as can be seen from Eq.~(\ref{eq:L}).  In 
Fig.~\ref{fig:LSP}, we show the region in which the LSP is the wino 
in the $r_*$-$|\mu|/m_{3/2}$ plane.  Here, we have set $m_{3/2} = 
100~{\rm TeV}$ and the phase of $\mu$ is chosen such that $|M_2|- |M_1|$ 
is minimized.  The value of $\tan\beta$ is chosen to realize the 
$125~{\rm GeV}$ Higgs boson.  (In the region where no solution for 
$\tan\beta$ exists, we have set $\tan\beta=1$.)

\subsection{Collider signals}
\label{subsec:collider}

\begin{figure}[t]
\begin{center}
  \includegraphics[clip,width=0.65\textwidth]{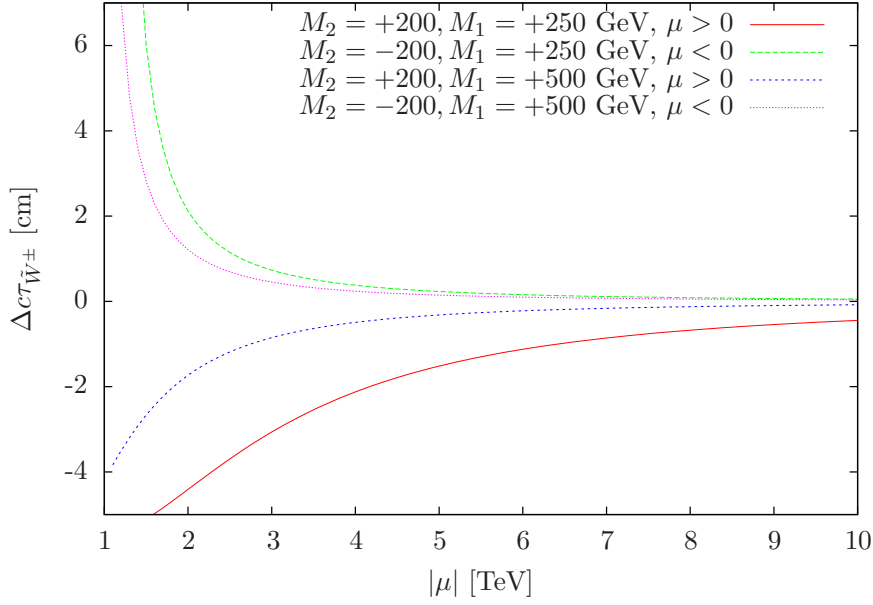}
\caption{$\varDelta c\tau_{\tilde W^\pm} = c\tau_{\tilde W^\pm} 
 - c\tau_{\tilde W^\pm}^{|\mu|\to\infty}$ as a function of $|\mu|$ for 
 various values of $M_1$ and $M_2$.  We have set $\sin(2\beta) = 1$.}
\label{fig:c_decay}
\end{center}
\end{figure}

As in any models in which the gaugino masses arise mainly from anomaly 
mediation and in which $\mu$ is sufficiently larger than the gaugino 
masses, the present model has the neutral wino LSP and the charged wino 
next-to-LSP nearly degenerate.  For $|\mu| = O(m_{3/2})$, the small 
mass difference between the charged and neutral winos is determined by 
the electromagnetic loop contribution: $\delta M = M_{\tilde{W}^\pm} 
- M_{\tilde{W}^0} \simeq 160~{\rm MeV}$.  This small mass splitting 
makes the decay length of the charged wino relatively long, $c\tau 
\approx O(10~{\rm cm})$, and make it potentially observable at the 
LHC~\cite{Ibe:2006de}.  Note, however, that in the present model, larger 
$r_*$ generically implies relatively small $|\mu|$ (though still of order 
$m_{3/2}$) and $\tan\beta$, because of the constraint from the wino LSP 
and the Higgs mass.  In this case, the tree-level contribution to the 
mass splitting can be important, which can be written as
\begin{equation} 
  (\delta M)^{\rm tree} \simeq 
    \frac{m_W^4 t_W^2 \sin^2(2\beta)}{|\mu|^2(|M_1|^2-|M_2|^2)} 
    \left\{ |M_2| + |M_1|\cos(\arg(M_1M_2\mu^2)) \right\},
\label{eq:delta-M_tree}
\end{equation}
where $t_W$ is the tangent of the Weinberg angle.  In fact, for small 
$|\mu|$ and/or $|M_1| \sim |M_2|$, the effect of this contribution on the 
decay width of the charged wino is sizable.  In Fig.~\ref{fig:c_decay}, 
we show the deviation of $c\tau_{\tilde W^\pm}$ from $c\tau_{\tilde 
W^\pm}^{|\mu|\to \infty}$.

As we will see later, a significant portion of the Spread Supersymmetry 
parameter region allows for production of gluinos at the LHC.  Once 
produced, the gluino decays through the exchange of heavy squarks. 
Therefore, the modes and rate of gluino decay can provide important 
information on the squark sector.  In the present model, the dominant 
decay modes are three-body decay $\tilde{g} \to q\bar{q}\chi$, 
and the two-body decay process $\tilde{g} \to g\chi$ is strongly 
suppressed~\cite{Toharia:2005gm,Gambino:2005eh}.  This therefore 
provides a good test of the model~\cite{Sato:2012xf}.

Because of $r_* \simgt O(1)$, the present model can have much heavier 
scalar particles than the conventional anomaly mediation model.  Such 
heavy scalars result in a long-lived gluino, whose decay length is 
approximately given by
\begin{equation}
  c\tau_{\tilde{g}} = O(1~{\rm cm})\, 
    \left( \frac{M_{\tilde{g}}}{1~{\rm TeV}} \right)^{-5} 
    \left( \frac{\tilde{m}}{1000~{\rm TeV}} \right)^4.
\label{eq:ctau-gluino}
\end{equation}
This implies that $r_*$ larger than $\approx O(10)$ may lead to distinct 
long-lived gluino signatures.  Together with the $\tilde{W}^\pm$ track 
arising from the gluino decay, such a long-lived gluino may allow us 
to extract various information, such as the lifetimes of $\tilde{W}^\pm$ 
and the gluino as well as the masses of the wino, bino and gluino.  An 
interesting possibility is that the long-lived gluino may ``carry'' the 
$\tilde{W}^\pm$ track from the collision point to the transition radiation 
tracker, facilitating the measurement of the $\tilde{W}^\pm$ track.

In Table~\ref{tb:gluino}, we compile the current constraints on the gluino 
mass for various gluino decay lengths.  While the accurate constraints 
depend on the details of the mass spectrum and decay patterns of the 
MSSM particles, and stopped gluino and $R$-hadron search are subject to 
different theoretical uncertainties, the lower bound on $M_{\tilde{g}}$ 
is roughly in the range of $\approx 1~{\rm TeV}$.
\begin{table}
\caption{Current lower bounds on the gluino mass.}
\label{tb:gluino}
\begin{tabular}{|c|c|c|} \hline
  $c\tau_{\tilde{g}}$ & Bound on $M_{\tilde{g}}$ [GeV] & References \\
\hline\hline
  Prompt& 700~--~1200 & \cite{multijets} (multi-jets + missing), 
    \cite{3b} ($b$ jets + missing), \ldots \\ \hline
  $30~{\rm m}$  & 500 & \cite{Chatrchyan:2012yg} (stopped gluino) \\ \hline
  $300~{\rm m}$ & 600 & \cite{Chatrchyan:2012yg} (stopped gluino) \\ \hline
  $3000~{\rm m}$~--~$10^{12}~{\rm m}$ & 640 & 
    \cite{Chatrchyan:2012yg} (stopped gluino) \\ \hline
  $\gg 10~{\rm m}$ & 1000~--~1100 & \cite{Chatrchyan:2012sp} ($R$-hadron)
\\ \hline
\end{tabular}
\end{table}

Summarizing, Spread Supersymmetry with wino LSP has nearly degenerate 
neutral wino LSP and charged wino next-to-LSP, as in other models based 
on anomaly mediation, and the latter may be detectable as a charged 
track at the LHC.  In addition, unlike the conventional anomaly 
mediation model, Spread Supersymmetry naturally leads to a long-lived 
gluino because of rather heavy squarks of mass $m_{\tilde{q}} \approx 
r_* m_{3/2}$.  Therefore, in some decay chains, we have two long-lived 
particles:
\begin{equation}
  \tilde{g} \;\; \xrightarrow[{\rm long-lived}]{} \;\; 
    q\bar{q}(\tilde{W}^\pm \xrightarrow[O(10~{\rm cm})]{} \tilde{W}^0 \pi^\pm ),
\end{equation}
allowing for an extraction of information about the masses and lifetimes 
of these particles.  Note that this signature, although exotic, is a natural 
consequence of the model, and occurs quite generically.  The measurement 
of this type of gluino decay may also reveal information on the squark 
sector, especially the scale of the squark masses.

\subsection{Flavor and {\boldmath $CP$}}
\label{subsec:flavor}

The heavier scalar mass spectrum is favorable from the viewpoint of 
the supersymmetric flavor and $CP$ problems.  The ultraviolet physics 
at $M_*$ need not respect the SM flavor symmetry, in which case large 
flavor violating soft masses are expected.  To suppress low energy flavor 
violating processes, large $\tilde{m}$ is then required.  For example, 
suppressing $\varDelta F = 2$ mixing between the first and second 
generation quarks requires $\tilde{m} > O(10^3~{\rm TeV})$, and 
that for the first and third requires $\tilde{m} > O(10^2~{\rm 
TeV})$~\cite{Gabbiani:1996hi}.  With $r_* > 1$, the present model 
can have sufficiently heavy scalar masses to avoid constraints from 
low-energy flavor violating processes, even if they have maximal 
flavor violation.  On the other hand, because we expect $r_* < O(100)$, 
some deviations from the SM may be observed in future flavor experiments.

The gluino decay is sensitive to the squark masses, and its observation 
may provide information about the flavor violating structure of the squark 
masses.  The gluino decays into two quarks and a lighter superparticle 
via a squark exchange.  If the squark masses have flavor-violating 
structure, the gluino will decay dominantly by the exchange of the lightest 
squark, leading to the quarks of the corresponding flavor.  Furthermore, 
flavor-violating decays, such as $\tilde{g} \rightarrow b\bar{s}\tilde{\chi}, 
t\bar{c}\tilde{\chi}$, are also expected to occur.  By using heavy 
flavor tagging techniques, such a ``flavorful'' gluino may be identified. 
Detailed studies of the flavor of the quarks from gluino decays, therefore, 
can provide important information on the size of flavor violation in 
the squark sector.

We finally discuss possible signals from electric dipole moments (EDMs). 
New physics beyond the standard model with $CP$ violation naturally provides 
large EDMs of an electron, a neutron, and so on.  In the present model, 
one-loop contributions to the EDMs from heavy scalars are suppressed 
unless $r_*$ is small, of $O(1)$.  However, the model predicts relatively 
small $\mu$, and its phase is generically expected to be of $O(1)$. 
In this case, two-loop diagrams without scalars can give significant 
contributions~\cite{ArkaniHamed:2004yi}; for example, the electron EDM 
is given by
\begin{equation}
  d_e \simeq 3 \times 10^{-29}~e~{\rm cm} \times 
    \sin(2\beta)\, \sin(\arg(M_2\mu))\, 
    \left( \frac{|\mu|}{10~{\rm TeV}} \right)^{-1} 
    \left( \frac{M_{\tilde W}}{200~{\rm GeV}} \right)^{-1} 
    f(m_h^2/M_{\tilde W}^2),
\label{eq:d_e}
\end{equation}
where $f(x) = 1 - \ln(x)/2 +(5/3 - \ln(x))x/12 + \cdots$.

The current constraint on the electron EDM is $d_e < 1.05 \times 
10^{-27}~e~{\rm cm}$ at $90\%$C.L.~\cite{Hudson:2011zz}, which still 
does not explore an interesting region of the model.  However, planned 
EDM experiments are expected to have a few orders of magnitude improved 
sensitivity~\cite{Hudson:2011zz,Vutha:2009ux}, reaching the level 
of $d_e \sim 10^{-31}~e~{\rm cm}$, or even smaller.  These experiments 
will then be a good probe of the model.

\section{Astrophysics and Cosmology}
\label{sec:cosmo}

In this section, we discuss astrophysical and cosmological aspects of 
Spread Supersymmetry with wino LSP.  We find that sizable $r_*$ plays 
an important role.

\subsection{Wino relic abundance}
\label{subsec:abundance}

In the present model, the gravitino mass is larger than $O(10~{\rm TeV})$. 
Such a heavy gravitino decays before BBN, so the model does not suffer 
from the BBN constraints~\cite{Kawasaki:2008qe}.  On the other hand, 
decays of gravitinos produced in the early universe lead to additional 
wino abundance after wino freeze-out.  Therefore, too large primordial 
gravitino abundance may result in overabundance of the wino LSP.

There are two sources for the relic abundance of the wino LSP:\ the 
thermal relic contribution and the non-thermal contribution from the 
gravitino decay.  The total wino relic abundance can thus be given by 
the sum of these two%
\footnote{If the gravitino mass is large and the wino mass small, then 
 late-time annihilation after the gravitino decay can be effective, 
 making the final wino abundance smaller.  If the gravitino mass is 
 $500~{\rm TeV}$ ($2000~{\rm TeV}$) and the wino mass $100~{\rm GeV}$ 
 ($500~{\rm GeV}$), for example, this leads to a deviation from the 
 simple sum in Eq.~(\ref{eq:Omega_wino}) by about $10\%$.  In the rest 
 of the paper, we ignore this effect for simplicity since it is not 
 significant in relevant parameter space.}
\begin{equation}
  \Omega_{\tilde{W}} = \Omega_{\tilde{W}}^{\rm thermal} 
    + \Omega_{\tilde{W}}^{\rm non-thermal}.
\label{eq:Omega_wino}
\end{equation}
The thermal relic abundance is given by
\begin{equation}
  \Omega_{\tilde{W}}^{\rm thermal} h^2 \simeq 2 \times 10^{-4} 
    \left( \frac{M_{\tilde{W}}}{100~{\rm GeV}} \right)^2
\label{eq:Omega-thermal}
\end{equation}
without including the Sommerfeld effect, which reduces 
$\Omega_{\tilde{W}}^{\rm thermal}$ from this expression for a heavy 
wino, $M_{\tilde{W}} \simgt {\rm TeV}$~\cite{Hisano:2005ec}.  The 
non-thermal wino abundance is related to the primordial gravitino 
abundance before the decay, $\Omega_{3/2}$, by
\begin{equation}
  \Omega_{\tilde{W}}^{\rm non-thermal} 
  = \frac{M_{\tilde{W}}}{m_{3/2}} \Omega_{3/2},
\label{eq:Omega-nonthermal}
\end{equation}
so we need to know $\Omega_{3/2}$ to obtain the final wino abundance, 
$\Omega_{\tilde{W}}$.

There are two main origins for the primordial gravitino abundance, 
$\Omega_{3/2}$.  One comes from thermal scatterings of the MSSM particles 
at the reheating era, which becomes important if the reheating temperature 
$T_R$ is high, e.g., $\simgt 10^8~{\rm GeV}$.%
\footnote{We define the reheating temperature by
 $$
 T_R \equiv \left[ \frac{90}{\pi^2 g_*(T_R)} 
 \Gamma_{\rm inf}^2 M_{\rm Pl}^2 \right]^{1/4},
 $$
 where $\Gamma_{\rm inf}$ is the decay rate of the inflaton field and 
 $g_*$ is the number of effective massless degrees of freedom.}
In this case (more specifically if $T_R \simgt O(10\,\tilde{m})$), 
the contribution depends almost only on $T_R$ and is well fitted 
by~\cite{Pradler:2006hh}
\begin{equation}
  \Omega_{3/2}^{\rm UV} h^2 \simeq 3.9\, 
    \left( \frac{T_R}{10^9~{\rm GeV}} \right) 
    \left( \frac{m_{3/2}}{100~{\rm TeV}} \right).
\label{eq:uv}
\end{equation}
(This contribution, however, suffers from some theoretical uncertainties; 
for example, the procedure of Ref.~\cite{Rychkov:2007uq} gives about 
two times larger abundance than the one in Eq.~(\ref{eq:uv}).)  Another 
one comes from the freeze-in contribution.  This contribution arises if 
$T_R$ is larger than the scalar masses $\tilde{m}$, and depends almost 
only on $\tilde{m}$:
\begin{equation}
  \Omega_{3/2}^{\rm freeze-in} h^2 
  \simeq 10^{-2}\!\!\! \sum_{i:\, {\rm thermalized}}\!\! 
    d_i\, \left( \frac{\tilde{m}_i}{1000~{\rm TeV}} \right)^3 
    \left( \frac{100~{\rm TeV}}{m_{3/2}} \right),
\label{eq:freeze-in}
\end{equation}
where $d_i$ is the degrees of freedom of superparticle $i$ of mass 
$\tilde{m}_i \sim \tilde{m}$.  This large freeze-in contribution 
of Eq.~(\ref{eq:freeze-in}) is a characteristic feature of Spread 
Supersymmetry---values of the cutoff scale $M_*$ smaller than 
$M_{\rm Pl}$ enhances the couplings between the scalar particles 
and the Goldstino components of the gravitino, strongly enhancing 
gravitino production.

The total primordial gravitino abundance is given by the sum of the above 
two contributions:
\begin{equation}
  \Omega_{3/2} = \Omega_{3/2}^{\rm UV} + \Omega_{3/2}^{\rm freeze-in},
\label{eq:Omega_32}
\end{equation}
which determines $\Omega_{\tilde{W}}^{\rm non-thermal}$ through 
Eq.~(\ref{eq:Omega-nonthermal}).  The relic wino LSP abundance is then given 
by adding the thermal relic contribution, in Eq.~(\ref{eq:Omega-thermal}).

\subsection{Detecting wino dark matter}
\label{subsec:detect}

\subsubsection*{Cosmic ray signals}

We now discuss the observability of relic wino LSPs.  Since the wino 
annihilation cross section is relatively large, it can potentially be 
probed by many processes, such as effects on BBN, distortion of cosmic 
microwave background (CMB), and production of cosmic rays, even if the wino 
may not comprise all of the dark matter.  In particular, cosmic-ray photon 
observation by the Fermi collaboration provides a significant constraint 
on relic winos.

\begin{figure}[t]
\begin{center}
  \includegraphics[clip,width=0.65\textwidth]{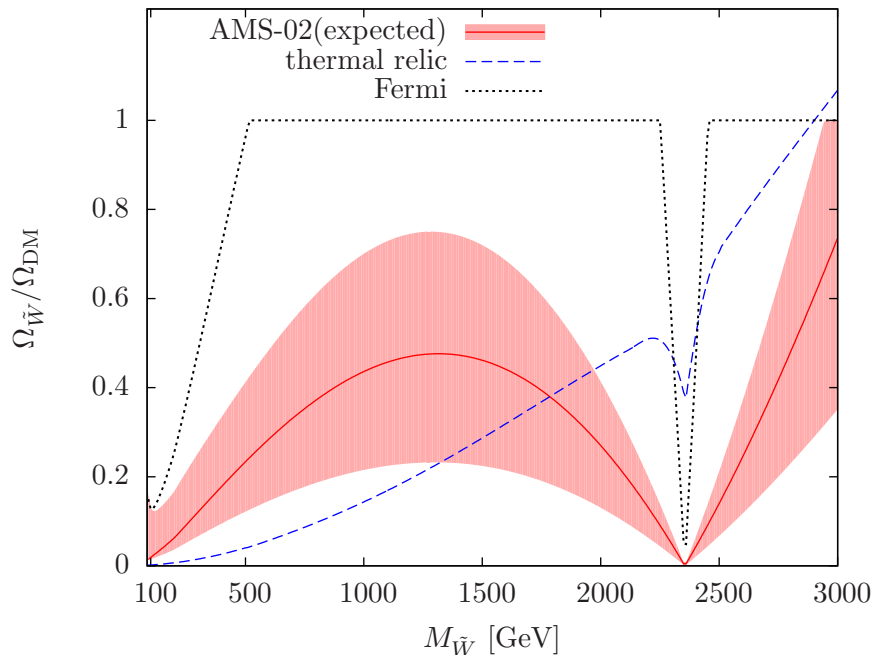}
\caption{Current and projected upper bounds on $\Omega_{\tilde{W}} / 
 \Omega_{\rm DM}$ as a function of the wino mass, $M_{\tilde{W}}$, 
 from the Fermi and AMS-02 cosmic ray experiments.  The red shaded 
 region shows the current uncertainty coming from the propagation and 
 dark matter halo models.  After the AMS-02 experiment, the uncertainty 
 from the cosmic ray propagation will be reduced.}
\label{fig:Fermi_cons}
\end{center}
\end{figure}

To discuss constraints from indirect detection experiments, including the 
Fermi observation, it is convenient to define the effective cross section
\begin{equation}
  \langle \sigma v \rangle_{\rm eff} 
  = \left( \frac{\Omega_{\tilde{W}}}{\Omega_{\rm DM}} \right)^2 
    \langle \sigma v \rangle_{\tilde{W}},
\label{eq:sigmav_eff}
\end{equation}
where $\langle \sigma v \rangle_{\tilde{W}}$ is the wino annihilation 
cross section (times velocity).  This is the quantity to be compared with 
the dark matter annihilation cross section in the usual indirect-detection 
exclusion plots (which assume $\Omega_{\tilde{W}} = \Omega_{\rm DM}$). 
The strongest current constraint comes from the Fermi gamma ray search of 
Milky way satellites~\cite{Ackermann:2011wa}.  In Fig.~\ref{fig:Fermi_cons}, 
we plot the upper bound on $\Omega_{\tilde{W}}/\Omega_{\rm DM}$ coming 
from this constraint as a function of the wino mass.  For comparison 
we show the relic abundance from thermal freeze-out.

Let us discuss the prospect from the future AMS-02 antiproton search, 
which may provide a very powerful probe for wino dark matter.  
Wino dark matter annihilations induce high energy antiprotons.  However, 
the flux of antiprotons on the top of the atmosphere strongly depends on 
the assumptions on the dark matter halo profile and propagation model 
for the antiproton cosmic rays.  Especially, uncertainties from the 
propagation models are huge and become a factor of 100~\cite{Evoli:2011id}. 
Actually, depending on the propagation parameters, the current antiproton 
measurement by the PAMELA Collaboration~\cite{Adriani:2010rc} can give 
a constraint stronger than that of Fermi.  The AMS-02 experiment has 
great advantages not only for measurements of antiprotons but also for 
other secondary-to-primary ratios such as boron-to-carbon (B/C).  High 
precision measurements of such quantities allow the propagation parameters 
to be estimated with higher accuracy, drastically reducing the astrophysical 
uncertainties of the antiproton flux~\cite{Pato:2010ih}. 
Hence, AMS-02 will be one of the strongest probes of Spread Supersymmetry. 

To estimate the sensitivity of the AMS-02 antiproton search, 
we have used the programs {\tt DRAGON}~\cite{Evoli:2008dv} and 
{\tt DarkSUSY}~\cite{Gondolo:2004sc}, to calculate the antiproton 
fluxes from astrophysical backgrounds and dark matter annihilations. 
We adopt the value of the acceptance and systematic errors of 
Ref.~\cite{Malinin:2004pw}.  Here we assume the systematic errors 
come from residual backgrounds, whose rate is $1~\mbox{--}10$~\% of 
the antiproton signals.  For the astrophysical background flux, we 
use the propagation model KRA of Ref.~\cite{Evoli:2011id}, and we simply 
assume that uncertainties of the background can be controlled with 
$\delta z_t = 1~{\rm kpc}$, where $z_t$ is the vertical size of the 
diffusion zone.  This size of the uncertainty will be reasonable after 
precise measurement of AMS-02~\cite{Pato:2010ih}.  Since the propagation 
parameters and the dark matter halo model are not determined well so 
far, we study some combinations of propagation models (KRA as well as 
MIN, MED and MAX models in Ref.~\cite{Delahaye:2007fr}) and dark matter 
profiles (NFW and isothermal).  We set the local dark matter energy 
density $\rho_{\odot}=0.4~{\rm GeV\cdot cm^{-3}}$.  Under the above 
simplified assumptions, we estimate the signal strength to detect 
deviation from the background at 95~\% C.L..  In Fig.~\ref{fig:Fermi_cons}, 
we show the sensitivity from the AMS-02 experiment.  The solid red line 
represents the case of KRA+isothermal and the shaded region shows the 
variation from propagation and dark matter halo models.  Note that 
this uncertainty mainly comes from ignorance of the underlying propagation 
model, which may be reduced after the experiment.

The current and projected bounds are clearly important especially when 
the non-thermal contribution in Eq.~(\ref{eq:Omega_wino}) dominates for 
low mass wino.  For heavier winos ($\simgt 2~{\rm TeV}$), AMS-02 may 
have the potential to probe wino dark matter, even if the relic abundance
is purely thermal.  This is because resonant process enhances the wino 
annihilation cross section~\cite{Hisano:2004ds} and heavier winos emit 
more high energy antiprotons, yielding a signal that is easier to discriminate 
from the astrophysical background.  Note that the results here are obtained 
under simplified assumptions; in reality, we would need to consider more 
detailed factors, such as breakdown of power-law primary proton injection, 
and perform more serious estimates of experimental systematic errors. 
Nevertheless, these results show that AMS-02 is likely to provide a very 
powerful probe of Spread Supersymmetry.

\subsubsection*{Direct detection}

Direct detection of wino dark matter is challenging.  The tree-level 
spin-independent dark matter-nucleon cross section is approximately 
given by
\begin{equation}
  \sigma_{\rm SI} \simeq (0.6-2) \times 10^{-46}~{\rm cm^2}\, 
    \sin^2(2\beta) \left( \frac{|\mu|}{5~{\rm TeV}} \right)^{-2} 
    \left( \cos(\arg(M_2\mu)) + \left|\frac{M_2}{\mu}\right| \right)^2.
\label{eq:sigma_SI}
\end{equation}
In the case that the $CP$-violating phase is nearly maximal in the relevant 
vertex, $\arg(M_2\mu) \simeq \pi/2$, $\sigma_{\rm SI}$ is highly suppressed 
and may be dominated by the second term or loop-induced contributions. 
For $\mu \sim 10~{\rm TeV}$, tree-level and loop-induced contributions 
are comparable and detailed calculation is required for a precise 
determination of the cross section.  When $|\mu| \gg 10~{\rm TeV}$, 
the cross section is dominated by the loop contribution and 
$\sigma_{\rm SI} \simeq 10^{-47}~{\rm cm}^2$.

The current constraint by the XENON100 experiment is~\cite{Aprile:2012nq}
\begin{equation}
  \sigma_{\rm SI} \simlt 1 \times 10^{-44}~{\rm cm}^2 
    \left( \frac{M_{\tilde{W}}}{500~{\rm GeV}} \right) 
    \left( \frac{\Omega_{\rm DM}}{\Omega_{\tilde{W}}} \right),
\label{eq:ZENON100}
\end{equation}
which does not reach the relevant parameter region even for 
$\Omega_{\tilde{W}} = \Omega_{\rm DM}$.

On the other hand, the projected sensitivity of XENON1T is~\cite{Aprile:2012zx}
\begin{equation}
  \sigma_{\rm SI} \simeq 1\times 10^{-46}~{\rm cm}^2 
    \left( \frac{M_{\tilde{W}}}{500~{\rm GeV}} \right) 
    \left( \frac{\Omega_{\rm DM}}{\Omega_{\tilde{W}}} \right),
\label{eq:ZENON1T}
\end{equation}
which is comparable with the cross section in Eq.~(\ref{eq:sigma_SI}) for 
$\Omega_{\tilde{W}} = \Omega_{\rm DM}$.  As can be seen in Fig.~\ref{fig:LSP}, 
for $r_* > 10$ a wino LSP requires $|\mu| < m_{3/2}$.  Therefore, future 
direct dark matter search experiments might detect the wino LSP in the 
optimistic case that it is the dominant component of dark matter.

\section{Results and Implications}
\label{sec:results}

We now investigate the parameter space of Spread Supersymmetry with wino 
LSP, based on the discussions so far.  We summarize the current status of 
the model and discuss future prospects for discovery.

\subsection{The current status}
\label{subsec:current}

The phenomenology of the model significantly depends on the scale of scalars 
$\tilde{m}$, the gravitino mass $m_{3/2}$, the size and phase of the $\mu$ 
parameter $|\mu| = O(m_{3/2})$ and $\arg(\mu)$, and $\tan\beta$.  The 
first two parameters can be traded with two real numbers $\sqrt{F_X}$ 
and $M_*$, and the next two with a complex parameter $L$ in Eq.~(\ref{eq:L}); 
the value of $\tan\beta$ can be determined by fixing $m_h = 125~{\rm GeV}$.
We therefore discuss parameter space of the model in terms of
\begin{equation}
  \sqrt{F_X}\, (= \sqrt{3}\, m_{3/2} M_{\rm Pl}),
\qquad
  M_*\, (= \sqrt{3}\, r_*^{-1} M_{\rm Pl}),
\qquad
  L = {\rm Re} L + i\, {\rm Im} L.
\end{equation}
In particular, the value of $L$ affects the ratios of the gaugino masses 
and has significant impacts on implications of the model.  The cosmology 
also depends on the reheating temperature $T_R$, so we have 5~parameters 
in total.

\begin{figure}[]
\begin{center}
  \includegraphics[clip,width=0.97\textwidth]{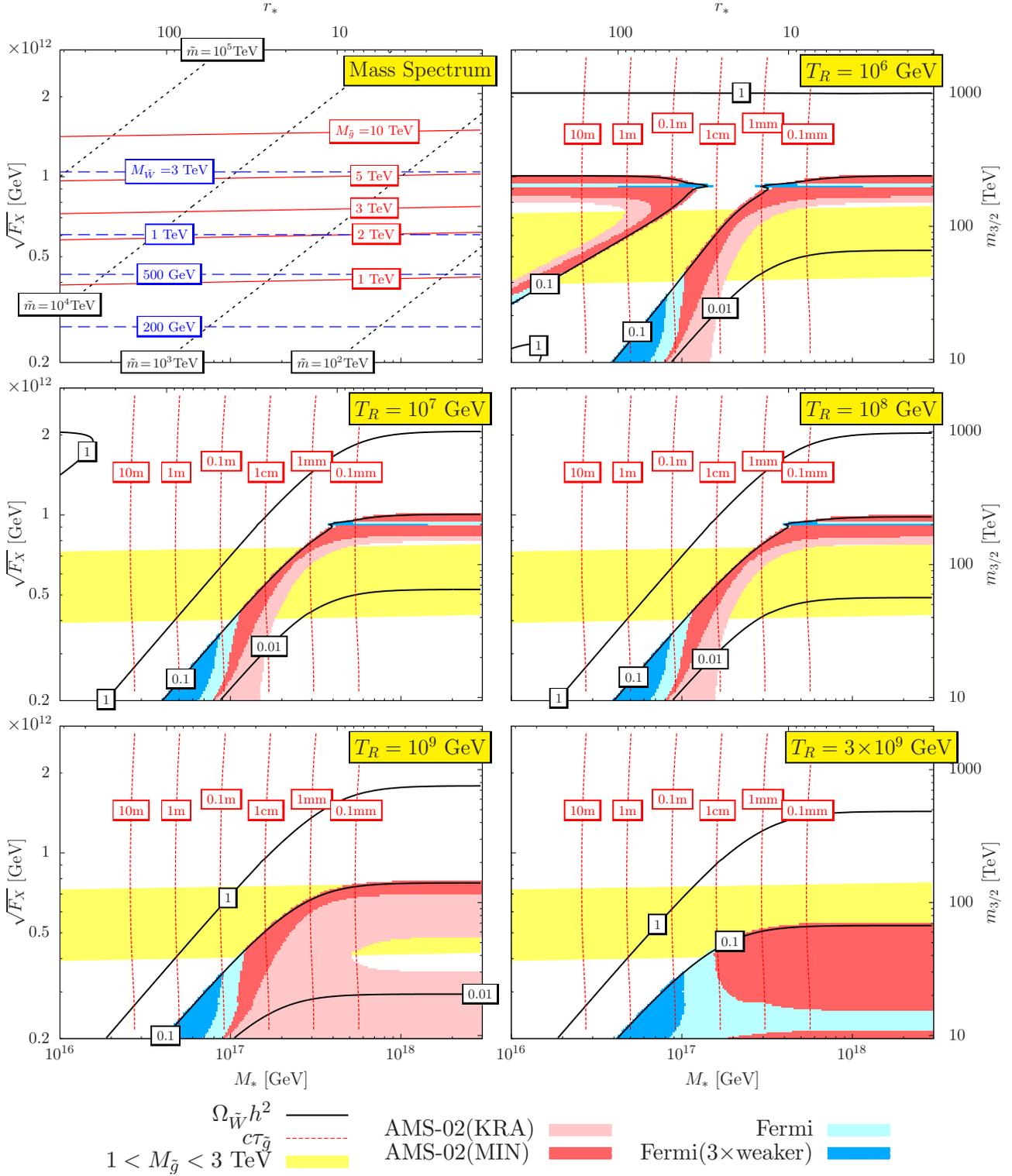}
\caption{Contours of the gluino decay length $c\tau_{\tilde{g}}$ and 
 the wino relic abundance $\Omega_{\tilde{W}} h^2$, as well as the 
 constraint from the Fermi photon observation and future prospect for 
 the AMS-02 antiproton search, are shown in the $M_*$-$\sqrt{F_X}$ (or 
 $r_*$-$m_{3/2}$) plane for various values of the reheating temperature 
 $T_R$.  Contours of the gluino and wino masses $M_{\tilde{g},\tilde{W}}$ 
 and the degenerate squark mass $\tilde{m}$ are also shown in the top 
 left panel.  The value of $L$ has been chosen such that $M_{\tilde{W}}$ 
 is maximized, keeping the wino LSP; numerically, $L \simeq 3 m_{3/2}$.}
\label{fig:max}
\end{center}
\end{figure}

\begin{figure}[]
\begin{center}
  \includegraphics[clip,width=0.97\textwidth]{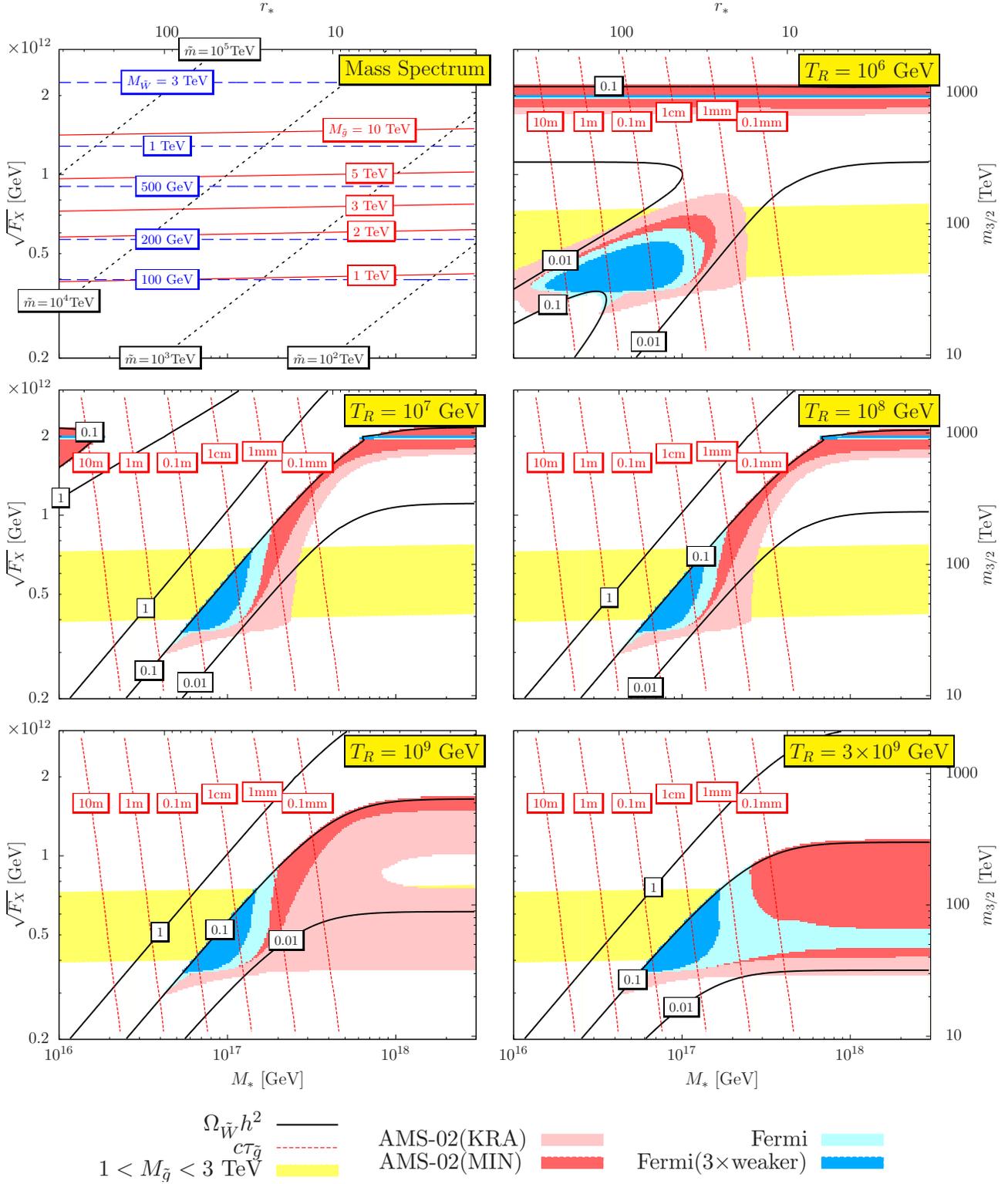}
\caption{Same as Fig.~\ref{fig:max} except that $L = 0$ (which leads to 
 the purely anomaly mediated gaugino spectrum).}
\label{fig:zero}
\end{center}
\end{figure}

\begin{figure}[]
\begin{center}
  \includegraphics[clip,width=0.97\textwidth]{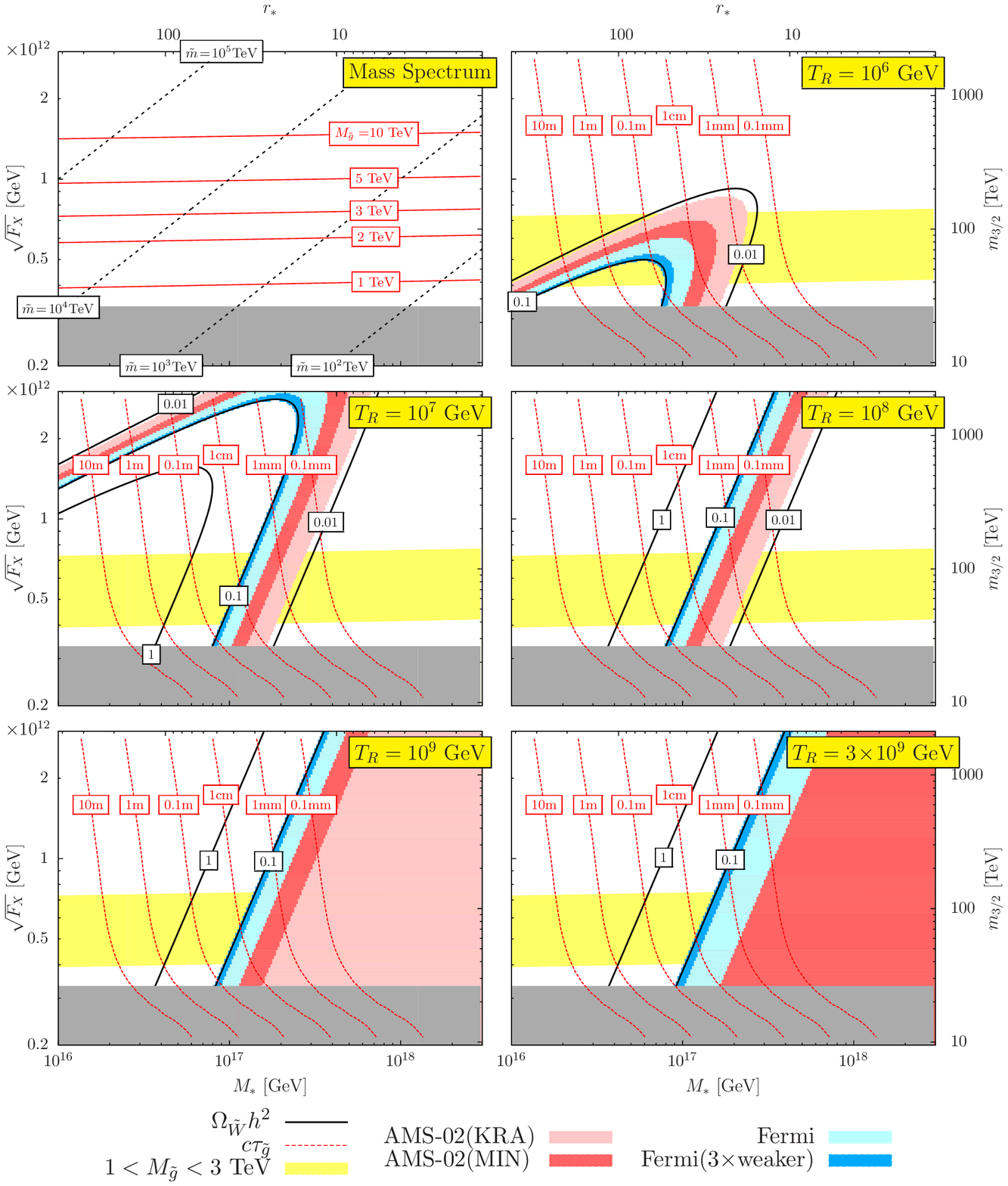}
\caption{The wino LSP abundance and some physical parameters are shown. 
 We choose $L$ so that $M_{2} = +300$ GeV.  The gray region is 
 the bino LSP.}
\label{fig:300}
\end{center}
\end{figure}

In Figs.~\ref{fig:max}~--~\ref{fig:300}, we plot selected physical 
quantities related to the dark matter and gluino properties in the 
$M_*$-$\sqrt{F_X}$ (or equivalently $r_*$-$m_{3/2}$) plane for several 
values for $T_R$, assuming that the scalar masses are degenerate with 
mass $\tilde{m}$.  (The effect of non-universality will be discussed 
later.)  Specifically, we plot the contours of the gluino decay length 
$c\tau_{\tilde{g}}$, the constraint from Fermi gamma ray search, and 
the wino relic abundance $\Omega_{\tilde{W}} h^2$; we also plot the 
contours of the gluino and wino masses $M_{\tilde{g},\tilde{W}}$ and 
the degenerate scalar mass $\tilde{m}$ in the top left panel of each 
figure.  Here, we have included in the Fermi constraint not only 
from Milky way satellite search but also from diffuse gamma ray 
search~\cite{Ackermann:2012qk}.  Since the cosmic-ray constraint 
potentially suffers from large astrophysical uncertainties, we also 
show the three times weaker constraint to be conservative.  We also 
show the future prospect for the AMS-02 antiproton search, using the 
isothermal dark matter profile.  Here we adopt KRA and MIN propagation, 
which provide medium and more conservative prospect, respectively. 
We have adopted a renormalization group-improved method for calculation 
of the gluino decay width~\cite{Gambino:2005eh}, and used {\tt 
micrOMEGAs~2.4}~\cite{Belanger:2010gh} in some parameter regions.

To see general features in the Spread Supersymmetry parameter space, in 
Figs.~\ref{fig:max} and \ref{fig:zero} we have chosen two representative 
values of $L$: $L \simeq 3 m_{3/2}$, which ``maximizes'' the wino 
mass keeping the wino LSP, i.e.\ $M_{\tilde{W}}$ is only slightly 
smaller than $M_{\tilde{B}}$ (Fig.~\ref{fig:max}) and $L = 0$, which 
corresponds to the case of a pure anomaly-mediated gaugino spectrum 
(Fig.~\ref{fig:zero}).  In a sense, these two cases represent two 
opposite ends of phenomenology that can be realized in the present model, 
corresponding to the cases with a small $M_{\tilde{g}}/M_{\tilde{W}}$ 
ratio (Fig.~\ref{fig:max}) and large $M_{\tilde{g}}/M_{\tilde{W}}$ ratio 
(Fig.~\ref{fig:zero}).  (A true extreme case, however, can occur when 
$L \simeq -m_{3/2}$, in which a cancellation of the anomaly mediated and 
loop contributions in the wino mass can make $M_{\tilde{g}}/M_{\tilde{W}}$ 
really large).

In both figures, we can see the following general trends.  If $T_R$ is 
sufficiently lower than $\tilde{m}$, e.g.\ as in the upper-left corner 
of $T_R = 10^6~{\rm GeV}$ panels, then the relic wino abundance is 
controlled only by the thermal freeze-out abundance; hence the contours 
of constant $\Omega_{\tilde{W}} h^2$ are horizontal.  Once $T_R$ becomes 
comparable or larger than $\tilde{m}$, however, the freeze-in contribution 
in Eq.~(\ref{eq:freeze-in}) can become important.%
\footnote{In the calculation of the gravitino abundance in the figures, 
 we have adopted $\rho_{\rm inf} \gg \rho_{\rm rad}$ as the initial 
 condition, where $\rho_{\rm inf}$ and $\rho_{\rm rad}$ are the inflaton 
 and radiation energy densities.  The reason why the freeze-in contribution 
 is relevant even for $T_R$ slightly smaller than $\tilde{m}$, then, is 
 that there are nonzero contributions from scalars heavier than $T_R$, 
 which is Boltzmann suppressed after the reheating, and that there is 
 a residuum from the era before the reheating, when $T \gg T_R$.}
In particular, this is more effective at larger $\tilde{m}$, i.e.\ smaller 
$M_*$ ($=$ larger $r_*$), bending the constant $\Omega_{\tilde{W}} h^2$ 
contours downward toward the left (as can be seen clearly in the plots 
with $T_R \geq 10^7~{\rm GeV}$).  This effect, therefore, prefers lower 
superparticle masses for small $M_*/M_{\rm Pl} \simlt O(0.1)$.  For larger 
$M_*$, the contours of constant $\Omega_{\tilde{W}} h^2$ are horizontal 
(i.e.\ do not depend on the scalar masses), which, however, become lower 
for large $T_R$ because of the contribution from the decay of the gravitino 
produced at the reheating, Eq.~(\ref{eq:uv}).

The condition from the relic wino abundance, therefore, provides upper 
bounds on the superparticle masses as a function of $M_*$ and $T_R$ for 
a given $L$.  For example, in the case of Fig.~\ref{fig:max} ($ L \simeq 
3 m_{3/2}$), the gluino mass is bounded very roughly as
\begin{equation}
  M_{\tilde{g}} \simlt 
  \left\{ \begin{array}{ll}
    1~{\rm TeV} \left( \frac{M_*}{10^{17}~{\rm GeV}} \right)^{3/2} 
      \left( \frac{\Omega_{\tilde{W}}}{\Omega_{\rm DM}} \right)^{1/2}\,\, &
    \mbox{for }\,\, M_* \simlt  10^{17}~{\rm GeV},
  \\
    {\rm min} \left\{ 5~{\rm TeV} 
      \left( \frac{\Omega_{\tilde{W}}}{\Omega_{\rm DM}} \right)^{1/2},\,
      2~{\rm TeV} \left( \frac{T_R}{3\times 10^9~{\rm GeV}} \right)^{-1} 
      \left( \frac{\Omega_{\tilde{W}}}{\Omega_{\rm DM}} \right) \right\}\,\, &
    \mbox{for }\,\, M_* \simgt 10^{17}~{\rm GeV}.
  \end{array} \right.
\label{eq:Mg_fig-5}
\end{equation}
This bound is much tighter than the naive anomaly mediated case of 
Fig.~\ref{fig:zero} ($ L = 0$):
\begin{equation}
  M_{\tilde{g}} \simlt 
  \left\{ \begin{array}{ll}
    2~{\rm TeV} \left( \frac{M_*}{10^{17}~{\rm GeV}} \right)^{3/2} 
      \left( \frac{\Omega_{\tilde{W}}}{\Omega_{\rm DM}} \right)^{1/2}\,\, &
    \mbox{for }\,\, M_* \simlt  10^{17}~{\rm GeV},
  \\
    {\rm min} \left\{ 20~{\rm TeV} 
      \left( \frac{\Omega_{\tilde{W}}}{\Omega_{\rm DM}} \right)^{1/2},\,
      7~{\rm TeV} \left( \frac{T_R}{3\times 10^9~{\rm GeV}} \right)^{-1} 
      \left( \frac{\Omega_{\tilde{W}}}{\Omega_{\rm DM}} \right) \right\}\,\, &
    \mbox{for }\,\, M_* \simgt  10^{17}~{\rm GeV}.
  \end{array} \right.
\label{eq:Mg_fig-6}
\end{equation}
We expect that a generic situation of the model is somewhere between these 
two cases, although larger values of the gluino mass are possible if $L$ 
is in the range $m_{3/2} \simlt |L| \simlt 3 m_{3/2}$ with $\arg(L) 
\simeq \pi$.

Given that flavor and $CP$ constraints require rather large $\tilde{m} 
> O(10^3~{\rm TeV})$ for generic scalar masses, we may naturally expect a 
somewhat small cutoff scale, $M_* \simlt \mbox{a few} \times 10^{17}~{\rm 
GeV}$.  Furthermore, if we require successful thermal leptogenesis, the 
reheating temperature must be high, $T_R \simgt 2 \times 10^9~{\rm 
GeV}$~\cite{Giudice:2003jh}.  These select the model to be in particular 
parameter regions.  While the regions start being constrained by the 
Fermi data, there are still significant regions remaining.  The gluino 
mass in these regions are less than a few TeV for generic values of $L$, 
so we may expect it to be within reach at the $13~{\rm TeV}$ run of the 
LHC.  Moreover, the decay length of the gluino in these regions is
\begin{equation}
  O(0.1~{\rm mm}) < c\tau_{\tilde{g}} < O(10~{\rm cm}),
\end{equation}
leading to the spectacular signal of a long-lived gluino decaying into 
a long-lived charged wino having $c\tau_{\tilde{W}^\pm} = O(10~{\rm cm})$.

In Fig.~\ref{fig:300}, we plot the contours of $c\tau_{\tilde{g}}$ and 
$\Omega_{\tilde{W}} h^2$, together with the Fermi constraint, fixing 
the wino mass rather than $L$: $M_2 = +300~{\rm GeV}$ (i.e.\ $|M_2| = 
300~{\rm GeV}$ with $\arg(M_2) = 0$).  We also plot the gluino mass 
$M_{\tilde{g}}$ and the scalar mass $\tilde{m}$ in the top left panel. 
We see that for smaller $M_*$, the gluino mass tends to be lighter and 
its decay length tends to be longer.
 
So far, we have assumed that the scalar masses are universal, but we 
generally expect this is not the case.  In order to include the effect 
of non-degeneracy of the scalar masses, we can consider that $\tilde{m}$ 
actually represents the ``effective scalar mass'' appearing in 
Eq.~(\ref{eq:freeze-in})
\begin{equation}
  \tilde{m}_{\rm eff} 
  = \left( \frac{\sum_i d_i \tilde{m}_i^3}{\sum_i d_i} \right)^{1/3},
\label{eq:tilde-m_eff}
\end{equation}
and that $M_*$ and $r_*$ in these figures are defined by this quantity:
\begin{equation}
  M_* \equiv \frac{F_X}{\tilde{m}_{\rm eff}},
\qquad
  r_* \equiv \frac{\sqrt{3}\, M_{\rm Pl}\, \tilde{m}_{\rm eff}}{F_X}.
\label{eq:eff-scales}
\end{equation}
The contours of the gluino decay length then represent not those of the 
true gluino decay length, $c\tau_{\tilde{g}}$, but of the ``rescaled'' 
gluino decay length
\begin{equation}
  c\tau_{\tilde{g},{\rm res}} = \tilde{m}_{\rm eff}^4 
    \left( \frac{1}{n_{\tilde{q}}} \sum_{\tilde{q}} 
    \frac{1}{m_{\tilde{q}}^4} \right) c\tau_{\tilde{g}}.
\label{eq:ctau_res}
\end{equation}
If there is a significant distribution in the squark masses, the 
true gluino decay length is then shorter than the values depicted 
in Figs.~\ref{fig:max}~--~\ref{fig:300}, which are obtained assuming 
universal scalar/squark masses.   A hierarchy in the squark spectrum 
could decouple the gluino lifetime from the cosmological wino abundance, 
with gluino decay dominated by the lightest squarks and gravitino 
freeze-in dominated by the heavier ones.

\subsection{Future prospects}
\label{subsec:future}

The present model can provide many phenomenological consequences, such as 
collider signals, dark matter signals, effects on precision physics, and 
so on.  Distinctive features of the model include scalar particles heavier 
than $m_{3/2}$ and Higgsinos with mass of order of $m_{3/2}$ or less. 
The heavy scalars provide the long-lived gluino, and the relatively 
lighter Higgsinos provide an enhanced possibility of indirectly detecting 
the Higgsino sector, compared to the conventional anomaly mediation model. 
In addition, the wino LSP may not comprise the whole dark matter, implying 
new possibilities for dark matter detection.  Here we discuss future 
prospects for these signatures.

The LHC has a great reach for the gluino.  The production cross section 
of gluinos is roughly $400$, $1$ and $0.01~{\rm fb}$ for $1$, $2$ and 
$3~{\rm TeV}$ gluino, respectively.  The signature there depends on the 
gluino lifetime.  For $c\tau_{\tilde{g}} \ll O(1~{\rm mm})$, the usual 
search for (missing energy~$+$~high $P_{\rm T}$ jets) is effective. 
For $\sqrt{s} = 14~{\rm TeV}$ and very heavy squarks, the LHC has 
a discovery reach of $m_{\tilde{g}}$ up to about $2.0~{\rm TeV}$ 
($2.3~{\rm TeV}$) with an integrated luminosity of $300~{\rm fb}^{-1}$ 
($3000~{\rm fb}^{-1}$)~\cite{Baer:2012vr,high_lumi_atlas}.  When 
$c\tau_{\tilde{g}} \simgt O(0.1~{\rm mm})$, the displaced vertex 
from the gluino decay can be recognized.  Although the current ATLAS 
study~\cite{displaced_atlas} of the displaced tracks assumes a specific 
decay topology of $R$-parity violation with a muon, a similar detection 
technique should work here as well for $c\tau_{\tilde g} \simgt 
O(0.1~{\rm mm})$, since the present model provides similar signals 
via gluino decay, e.g.\ $\tilde{g} \rightarrow \tilde{B}qq \rightarrow 
\tilde{W}Wqq$.  Applications to different decay topologies, such as 
displaced vertex~$+$~electron, are also expected to work.  Note that 
the gluino lifetime is very sensitive to the squark mass.  Therefore, 
the discovery of gluinos with displaced vertices would have a significant 
impact on the cosmology of the model, since the squark mass plays 
a crucial role in determining the dark matter abundance.

Disappearing tracks of charged winos are also interesting signals.  As 
discussed before, in the present model, the decay length of the charged 
wino may differ from the prediction of the conventional anomaly mediation, 
because of the contribution from the Higgsino to the charged-neutral wino 
mass splitting.  If the LHC or a future linear collider can determine the 
precise decay length of the charged wino and/or mass splitting between 
the charged and neutral winos, it would be possible to explore the 
Higgsino sector through these measurements.

Other important probes of the model come from the smallness of the 
$\mu$ term, which enhances dark matter direct detection as well as EDM 
detection.  The sensitivities of experiments exploring these signals 
are expect to be drastically improved in the future, enough to probe 
the case with $|\mu| \simlt 10~{\rm TeV}$ as discussed before.

Even if the wino mass is less than about $1~{\rm TeV}$, non-thermal 
wino production can give a significant relic density, although not 
necessarily $\Omega_{\tilde{W}} = \Omega_{\rm DM}$, allowing the wino 
LSP to be probed via cosmic rays and the CMB.  In particular, searches 
for antiproton cosmic rays in AMS-02 is very powerful for detecting 
signals from wino annihilation.  Also, the observation of CMB via the 
Planck satellite gives significant information of the wino dark matter. 
In particular, the large cross section of the wino has a great impact 
on the process of recombination in the early universe, and this effect 
can be probed via detailed observation of the CMB.  The current and 
expected limits are given by~\cite{Galli:2009zc,Slatyer:2009yq}
\begin{equation}
  m_{\tilde W} \simlt
    \left( \frac{\Omega_{\tilde{W}}}{\Omega_{\rm DM}} \right)^{2/3} 
    \times \left\{ \begin{array}{cl}
      230 ~{\rm GeV} & ({\rm WMAP7}) \\
      460 ~{\rm GeV} & ({\rm Planck~forecast}) \\
      700~{\rm GeV} & ({\rm cosmic~variance~with~} \ell_{\max}=2500)
    \end{array} \right.
\label{eq:CMB}
\end{equation}
at 95 \% C.L..  Both experiments are ongoing and will release the data 
in the near future.  Observing $\Omega_{\tilde{W}} < \Omega_{\rm DM}$ 
in these experiments would provide substantial evidence of our model.

Finally, let us discuss another way to test the heavy scalar sector by 
observations of primordial background gravitational waves.  The scalar 
particles affect the expansion of the universe.  At temperatures 
near $\tilde{m}$, a sudden change of relativistic degrees of freedom 
$\varDelta g_* \sim 100$ is expected, affecting the propagation 
of background gravitational waves.  The next generation of gravitational 
wave experiments, such as the Deci-hertz Interferometer Gravitational 
Wave Observatory (DECIGO)~\cite{DECIGO} and Big Bang Observer 
(BBO)~\cite{BBO}, have a potential to test this effect if $\tilde{m} 
> O(1000~{\rm TeV})$ and the primordial gravitational wave has a large 
amplitude~\cite{Saito:2012bb}.

\section{Multiverse Interpretation}
\label{sec:multiverse}

While the theoretical structure of Spread Supersymmetry with $\tilde{W}$ 
LSP is extremely simple, there is a remarkable coincidence in the value 
that the key parameters must take for it to be realistic.  There are two 
key mass scales in the theory beyond those of the SM:\ the supersymmetry 
breaking scale $F_X$ that sets $m_{3/2}$, and the mediation scale $M_*$ 
that then determines $\tilde{m}$.  In principle these parameters could take 
values varying over many orders of magnitude.  In practice, we see from 
Figs.~\ref{fig:max} and \ref{fig:zero} that the region of interest is quite 
small; if $\sqrt{F_X} \simlt 2\times 10^{11}~{\rm GeV}$ the wino would have 
been discovered at LEP and if $\sqrt{F_X} \simgt 2 \times 10^{12}~{\rm GeV}$ 
there would be too much dark matter.  Similarly, $M_*$ must be within two 
orders of magnitude of the reduced Planck mass.  A third mass scale, $T_R$, 
is crucial for the cosmological abundance of dark matter.  This scale 
is varied over several orders of magnitude in the various panels of 
Figs.~\ref{fig:max} and \ref{fig:zero}, and is apparently bounded only 
by $T_R \simlt 10^{10}~{\rm GeV}$ to avoid too much dark matter from 
UV production of gravitinos.  However, to obtain a baryon asymmetry via 
thermal leptogenesis requires $T_R \simgt 2 \times 10^9~{\rm GeV}$, so 
this mass scale may be tightly constrained also.

The remarkable coincidence is that this small parameter region is precisely 
where the three independent relic wino production mechanisms (thermal 
LSP freeze-out, gravitino freeze-in, and UV gravitino production) yield 
comparable contributions to the dark matter abundance
\begin{eqnarray}
  (Ym)_{\tilde{W}}|_{\rm FO} &\sim& 10^{-10}~{\rm GeV} 
    \left( \frac{\sqrt{F_X}}{2 \times 10^{12}~{\rm GeV}} \right)^4 
    \left( \frac{2 \times 10^{18}~{\rm GeV}}{M_{\rm Pl}} \right)^3,
\label{eq:OmegaFO}\\
  (Ym)_{\tilde{W}}|_{\rm FI} &\sim& 10^{-10}~{\rm GeV} 
    \left( \frac{\sqrt{F_X}}{2 \times 10^{12}~{\rm GeV}} \right)^4 
    \left( \frac{3 \times 10^{17}~{\rm GeV}}{M_*} \right)^3,
\label{eq:OmegaFI}\\
  (Ym)_{\tilde{W}}|_{\rm UV} &\sim& 10^{-10}~{\rm GeV} 
    \left( \frac{T_R}{10^9~{\rm GeV}} \right) 
    \left( \frac{\sqrt{F_X}}{2 \times 10^{12}~{\rm GeV}} \right)^2 
    \left( \frac{2 \times 10^{18}~{\rm GeV}}{M_{\rm Pl}} \right)^2.
\label{eq:OmegaUV}
\end{eqnarray}
With squark masses of order $10^3~{\rm TeV}$ the fine-tuning in 
electroweak symmetry breaking is of order $1$ in $10^8$, suggesting 
that the weak scale is anthropically selected.  In a multiverse view, 
the small values of the cosmological constant and the weak scale can both 
be understood as a consequence of environmental selection; significantly 
larger values would have catastrophic consequences for large scale 
structure~\cite{Weinberg:1987dv} and for stable nuclei~\cite{Agrawal:1997gf}. 
Could the coincidence of three comparable contributions to LSP dark matter 
follow from the environmental selection of the relevant mass scales?

Recall that in any version of split supersymmetry the abundance of dark 
matter has been logically disconnected from the weak scale---there is 
no ``WIMP miracle.''  Instead we assume an anthropic requirement on the 
total energy density of dark matter, either $\rho > \rho_{\rm cat}$ or 
$\rho < \rho_{\rm cat}$, where $\rho_{\rm cat}$ is the energy density 
at some catastrophic boundary, and we consider both possibilities.

\begin{figure}[]
\begin{center}
  \includegraphics[clip,width=0.65\textwidth]{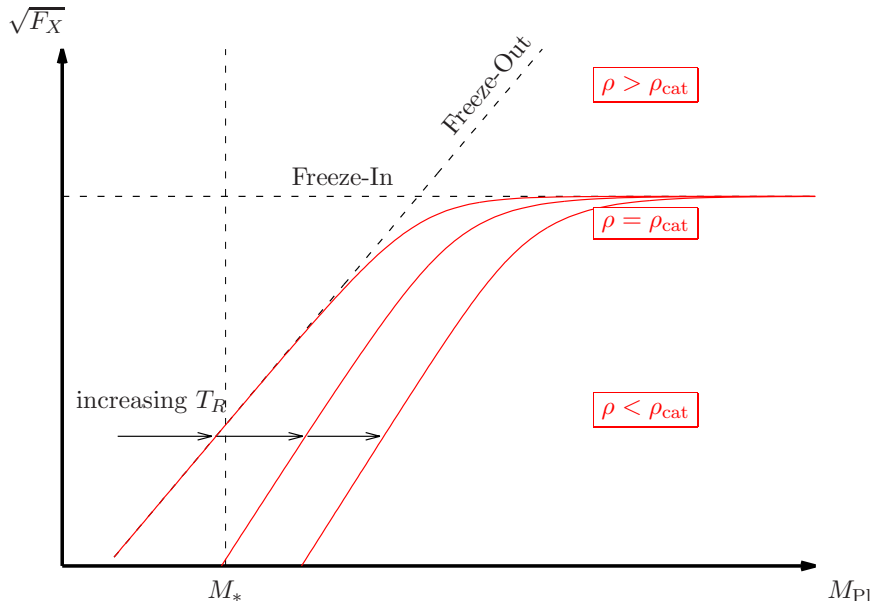}
\caption{Schematic picture of the catastrophic boundary for the dark 
 matter energy density.}
\label{fig:cat}
\end{center}
\end{figure}

We take $M_*$ to be the fundamental mass scale of the theory, and consider 
scanning of $M_{\rm Pl}$, $\sqrt{F_X}$ and $T_R$ over a wide range of 
values in the multiverse.  The red curves in Fig.~\ref{fig:cat} show 
the catastrophic boundary $\rho = \rho_{\rm cat}$ in the $(M_{\rm Pl}, 
\sqrt{F_X})$ plane for three values of $T_R$, assuming that $\rho_{\rm cat}$ 
is a constant, independent of these scanning parameters.  For simplicity 
we begin by fixing $T_R$ small enough that UV production of gravitinos 
is negligible, and add scanning of $T_R$ later, so for now we focus on 
the upper red curve of Fig.~\ref{fig:cat}.  For low values of $M_{\rm Pl}$ 
the boundary is determined by LSP freeze-out, while at higher values it 
is determined by gravitino freeze-in, which is independent of $M_{\rm Pl}$. 
Suppose that the environmental requirement is $\rho < \rho_{\rm cat}$ 
and that the multiverse distributions favor a large $\sqrt{F_X}$ 
and small $M_{\rm Pl}$ such that the most probable universes satisfying 
the environmental bound are located just below the kink in the 
catastrophic boundary.  This special location is precisely where 
$(Ym)_{\tilde{W}}|_{\rm FO}$ and $(Ym)_{\tilde{W}}|_{\rm FI}$ are 
comparable.  Alternatively, if the environmental requirement is 
$\rho > \rho_{\rm cat}$ then the multiverse distributions must favor 
small values of both $\sqrt{F_X}$ and $M_{\rm Pl}$ for the most probable 
observed universes to be close to the catastrophic boundary.  However, 
in this case there appears to be runaway behavior along the boundary. 
However, such a runaway is halted by $M_{\rm Pl}$ reaching $M_*$. 
Comparing Eqs.~(\ref{eq:OmegaFO}) and (\ref{eq:OmegaFI}), we see that 
this happens only an order of magnitude below the kink in the catastrophic 
boundary.  Hence this case also explains the coincidence of comparable 
freeze-in and freeze-out contributions to dark matter, but favors the 
freeze-out contribution slightly dominating.

We now add the scanning of $T_R$ to the above picture.  For $T_R 
< 10^8~{\rm GeV}$, $(Ym)_{\tilde{W}}|_{\rm UV}$ is sub-dominant so that 
the catastrophic boundary is essentially unaltered and is the upper red 
curve of Fig.~\ref{fig:cat}.  However for larger values of $T_R$, the 
catastrophic boundary at lower values of $M_{\rm Pl}$ is pushed down to 
lower $\sqrt{F_X}$ by UV production of gravitinos, as shown by the two 
lower red curves in Fig.~\ref{fig:cat}.  If the environmental requirement 
is $\rho > \rho_{\rm cat}$, it is not possible to understand the triple 
coincidence of freeze-out, freeze-in, and UV contributions---if the 
multiverse distribution favors small $T_R$ the UV contribution is negligible, 
while if it prefers high $T_R$ there is runaway behavior to high $T_R$ 
and low $\sqrt{F_X}$.  However, if the environmental requirement is 
$\rho < \rho_{\rm cat}$, the triple coincidence is easy to understand. 
If the multiverse prefers large $T_R$, but with a distribution that is 
not as strong as that for large $\sqrt{F_X}$ and small $M_{\rm Pl}$, 
then the most probable value of $T_R$ will be where it just starts to 
affect the catastrophic boundary with $T_R$ near $10^9~{\rm GeV}$.  If 
the catastrophic boundary is drawn as a surface in $(M_{\rm Pl}, \sqrt{F_X}, 
T_R)$ space, the triple coincidence occurs in universes that lie near the 
``tip of the cone'' of this surface~\cite{Hall:2007ja}.  With right-handed 
neutrino masses scanning with a distribution favoring large values, then 
an anthropic requirement of sufficient baryon asymmetry will force at 
least one right-handed neutrino mass to be at the $10^9~{\rm GeV}$ scale.

If the physics of the catastrophic boundary for dark matter depends 
on gravity, or on the expansion rate of the universe, then we expect 
$\rho_{\rm cat}$ to depend on $M_{\rm Pl}$, changing the location of 
the boundary in Fig.~\ref{fig:cat}.  However, providing the dependence 
on $M_{\rm Pl}$ is rather featureless, for example a simple power law, 
the boundary will simply appear rotated in Fig.~\ref{fig:cat}, and will 
still display the crucial kink.  If the probability distribution grows 
towards the kink then the understanding of the coincidences is preserved. 
For example, for $\rho < \rho_{\rm cat} \propto 1/M_{\rm Pl}^n$, with 
$n=1,2$, the coincidence results providing the strongest distribution 
favors large $\sqrt{F_X}$.

Finally, it has recently been shown that in a theory such as ours, with 
$|\mu| < \tilde{m} \ll M_*$, a somewhat special boundary condition on 
the top squark and up-type Higgs masses at the scale $M_*$ is required 
to avoid color breaking minima while allowing electroweak symmetry 
breaking~\cite{Arvanitaki:2012ps}.  However, if electroweak symmetry 
breaking and color conservation are viewed as environmental requirements 
on a multiverse, then this region of scalar masses will be selected 
environmentally in the multiverse.

\section{Conclusion}
\label{sec:concl}

Spread Supersymmetry with wino LSP is a particularly simple theory 
of Split Supersymmetry resulting from the supersymmetry breaking 
of Eq.~(\ref{eq:susybr}), leading to the typical spectrum shown in 
Fig.~\ref{fig:spread}.  The squarks are within about an order of magnitude 
of $10^3~{\rm TeV}$ and $\tan\beta$ is expected to be order unity, 
leading to a prediction for the mass of a SM-like Higgs boson shown 
in Fig.~\ref{fig:higgsmass}.  The recent discovery of a $125~{\rm GeV}$ 
SM-like Higgs boson strongly motivates further study of this theory.

Our key results are shown in Figs.~\ref{fig:max} and \ref{fig:zero} 
for two different values of $\mu$:\ these two figures represent opposite 
extremes of the expected phenomenology, with Fig.~\ref{fig:max} 
(\ref{fig:zero}) having a small (large) gluino to wino mass ratio. 
The solid black lines show our results for the cosmological wino 
abundance, the red dashed lines show our predictions for the gluino 
lifetime and the yellow bands show regions with the gluino mass in 
the $(1~\mbox{--}~3)~{\rm TeV}$, in reach of the LHC.  If the fundamental 
mass scale, $M_*$, that acts as a cutoff to the effective theory and 
is the mediation scale for supersymmetry breaking, is of order the 
Planck scale, then gluino decays are always prompt.  Also, in this 
case if $\Omega_{\tilde{W}} h^2 \simeq 0.1$ the gluinos are too heavy 
to discover at the LHC, unless $T_R$ is near its maximal value of 
order $10^9~{\rm GeV}$.

In Spread Supersymmetry, however, allowing for smaller values of $M_*$ 
opens up a region of parameter space with interesting experimental 
signatures.  Requiring that the wino abundance $\Omega_{\tilde{W}} h^2 
< 0.1$, we find $M_* > 2 \times 10^{16}~{\rm GeV}$ throughout the region, 
preserving a highly successful gauge coupling unification.  The resulting 
region of interest has some dependence on both $\mu$ and $T_R$.  Crucially, 
a large fraction of this region with $T_R > \tilde{m}$ has gluinos within 
reach of LHC.  Furthermore, provided the squark spectrum is not too 
hierarchical, in a large fraction of the region gluinos decay with 
displaced vertices.  Gluino decays also lead to tracks from long-lived 
charged winos and may give flavor violating signals.

Much of this allowed region has $\Omega_{\tilde{W}} h^2 > 0.01$, as 
expected in theories where dark matter is environmentally selected 
in the multiverse.  AMS-02 searches for cosmic ray antiprotons will 
provide an important probe of this region.

\section*{Acknowledgments}

S.S. thanks N. Nagata for valuable discussions. 
This work was supported in part by the Director, Office of Science, 
Office of High Energy and Nuclear Physics, 
of the US Department of Energy under Contracts DE-FG02-05ER41360 and DE-AC02-05CH11231, 
in part by the National Science Foundation under grants PHY-0855653 and PHY-1002399 
and in part by the EU ITN grant UNILHC 237920 (Unification in the LHC era). 
L.J.H. and Y.N. acknowledge the hospitality of the Aspen Center for Physics, which 
is supported by the National Science Foundation Grant No.\ PHY-1066293.

\end{document}